\documentclass[useAMS,usenatbib]{mn2e}

\usepackage{graphicx}

\def\Ha{H$_{\alpha}$}
\def\Msol{M$_{\odot}$}
\def\St{$\Sigma_{GAS}/\Sigma_{CRIT}$}
\def\Str{($\Sigma_{GAS}/\Sigma_{CRIT})_{thres}$}

\title{The Star Formation Rate in disk galaxies: thresholds and dependence on gas amount }

\author[S. Boissier et al.]{S. 
Boissier$^{1}$\thanks{E-mail:boissier@ociw.edu}, N. Prantzos$^{2}$, 
A. Boselli$^{3}$ and G. Gavazzi$^{4}$\\
$^{1}$Carnegie Observatories, 813 Santa Barbara Street, Pasadena, California 91101, USA \\
$^{2}$Institut d'Astrophysique de Paris, 98bis Bd. Arago, 75104 Paris, France \\
$^{3}$Laboratoire d'Astrophysique de Marseille, Traverse du Siphon, F-13376 Marseille Cedex 12, France \\
$^{4}$Universita degli Studi di Milano - Bicocca, Pizza dell'Ateneo Nuovo 1, 20126 Milano, Italy }

\begin{document}

\date{submitted july 2003}

\pagerange{\pageref{firstpage}--\pageref{lastpage}} \pubyear{2003}

\maketitle

\label{firstpage}

\begin{abstract}
We reassess the applicability of the Toomre criterion in
   galactic disks and we study the local star formation law in 16 disk
   galaxies for which abundance gradients are published.  The data we
   use consists of stellar light profiles, atomic and molecular gas
   (deduced from CO with a metallicity-dependent conversion factor),
   star formation rates (from H$_{\alpha}$ emissivities),
   metallicities, dispersion velocities and rotation curves. We show
   that the Toomre criterion applies successfully to the case of the
   Milky Way disk, but it has limited success with the data of our
   sample; depending on whether the stellar component is included or
   not in the stability analysis, we find average values for the
   threshold ratio of the gas surface density to the critical surface
   density in the range 0.5 to 0.7. We also test various star
   formation laws proposed in the literature, i.e.  either the simple
   Schmidt law or modifications of it, that take into account
   dynamical factors. We find only small differences among them as far
   as the overall fit to our data is concerned; in particular, we find
   that all three SF laws (with parameters derived from the fits to
   our data) match particularly well observations in the Milky Way
   disk. In all cases we find that the exponent $n$ of our best fit
   SFR has slightly higher values than in other recent works and we
   suggest several reasons that may cause that discrepancy.
\end{abstract}


\section{Introduction}

Star formation is the main driver of galactic evolution.
Despite four decades of intense observational and theoretical investigation
(see e.g. Elmegreen 2002 for a recent overview) our understanding of the
subject remains frustratingly poor. As a result, important questions
related e.g. to the putative threshold and to the rate of star formation
in galaxies have no clear answers yet.

In most (if not all) studies of galaxy evolution, empirical formulae are used
to describe the star formation. Such formulae are based on the original
suggestion by Schmidt (1959), namely that the star formation rate (SFR) is 
simply proportional to some power $n$ of the gas mass density $\rho$. 
In the case of disk galaxies and starbursts observations (Kennicutt
1998b) suggest that such a relation indeed holds when gas surface density
$\Sigma_{GAS}$ is used instead of the mass density and when quantities
are averaged over the whole optical disk; in that case,
the exponent $n$ is found to be close to 1.5. However, Kennicutt (1998b)
noted that other interpretations of the data are possible: in particular,
an equally good fit to the data is obtained when it is assumed that gas turns 
into stars within a dynamical timescale (taken to be the orbital timescale
at the optical radius).

Such ``global'' laws may be useful for some applications 
(like e.g. semi-analytical models of galaxy evolution) but for 
detailed models of galactic disks ``local'' SF laws are required. Several
such laws have been suggested either on observational
(e.g. Dopita \& Ryder, 1994) or theoretical grounds (e.g. dynamical
instabilities in spirals, Ohnisi 1975, Wyse \& Silk 1989). In a recent work,
Wong \& Blitz (2002) find that the simple Schmidt law is valid locally,
with $n$=1.1-1.7 (depending on whether extinction on the observed
H$_{\alpha}$ emissivity profiles of their disks is assumed to be uniform
or dependent on gas column density). They also find that a dynamically
modified Schmidt law
(i.e. by introducing the orbital timescale as Kennicutt 1998b)
gives also an acceptable fit to their data.

The question of  a threshold in  star formation in galactic 
disks has been observationally assessed by Kennicutt (1989), 
who found that star formation is strongly suppressed below 
$\Sigma_{GAS}\sim$5-10 M$_{\odot}$ pc$^{-2}$. The existence of 
such a threshold is indeed 
suggested by dynamical stability analysis of thin, gaseous, differentially
rotating disks (e.g. Toomre 1964, Quirk 1972). 
In a recent investigation, Martin and
Kennicutt (2001) found that the threshold gas density (measured at the
edge of the star forming disk) varies by at least an order of magnitude
among spiral galaxies, but the ratio of gas density to a critical density
(see Sect. 4 for its definition) is much more uniform. 
Martin \& Kennicutt (2001) also identified the limitations of 
their observational strategy: uncertainties associated with non axisymmetric
gas distributions and with the extrapolation of the molecular gas density
profile, from the last observed point to the edge of the star forming disk.
The latter issue was  properly studied by Wong \& Blitz (2002)
with the help of detailed molecular profiles, but for a limited sample of
spirals, particularly rich in molecular gas. 
Their conclusion is that the gravitational
stability criterion has only a limited application.

From the theoretical point of view, quite a lot of work has been done to 
include more physics in the analysis than the simple stability criterion 
(e.g. Toomre 1981, Elmegreen 1987, Romeo 1992, Wang \& Silk 1994 etc.)
The general conclusion is that the various physical factors
(stars, magnetic fields, turbulence) affect only slightly the stability
parameter, by a factor of order unity.

In a recent work Schaye (2002) reassesses the role of the gas velocity
dispersion in the application of the simple stability criterion. 
He suggests that the usual assumption of a constant dispersion over the disk
may not hold and he investigates the stability  of thin, gaseous, 
self-gravitating disks imbedded in dark halos and illuminated by UV radiation.
He finds that the drop in the velocity dispersion associated with the 
transition from a warm to a cold phase of the interstellar medium
(i.e. from $\sim$10$^4$ K to below 10$^3$ K)
causes the disk to become gravitationally unstable.
This analysis leads to prescriptions for evaluating threshold
surface densities as a function of metallicity, intensity of UV radiation,
and gaseous fraction. However, a major prediction of his model analysis
seems to be contradicted by observations. Indeed, he finds that at the
phase transition there is always a sharp increase in the molecular gas
fraction, whereas Martin \& Kennicutt (2001) 
find that at the thresholds of their
star forming disks the gas is primarily atomic (for the low gas surface
densities) or molecular (for the highest gas surface densities), i.e.
no clear sign of a phase transition.

In this work, we reassess the applicability of the Toomre criterion
in galactic disks and we study the local star formation law
by means of a homogeneous sample containing 16 
spirals, half of which belong to the Virgo cluster.
In Sect. 2 we present our data sample, which consists of profiles
of stars, atomic and molecular gas, star formation rates (from H$_{\alpha}$
emissivities), metallicities and  rotation curves. In Sect. 3 we present
briefly the theoretical background of the stability criterion, including
a modification suggested by Wang \& Silk (1994). We show that it applies
successfully to the case of the Milky Way disk (Sect. 3.2) but it has
limited success with the data of our sample. In Sect. 4 we test various
SF laws proposed in the literature, i.e.
either the simple Schmidt law or modifications of it, that take into
account dynamical factors. We find only small differences among them
as far as the overall fit to our data is concerned; in particular, we find 
that all three SF laws (with parameters derived from the fits to our data)
match particularly well observations in the Milky Way disk.
In all cases 
we find that the exponent $n$ of our best fit SFR has slightly higher values 
than  in other recent works and we suggest several reasons that may cause
that discrepancy.
Our results are summarised in Sect. 5.

\section{Observational data}
\label{secdata}

\subsection{The galaxy sample}

The galaxies of our sample were selected for a detailed study of
the star formation properties in nearby spirals. Our first requirement
was that abundance measurements in HII regions are available so that
a metallicity gradient is defined. The molecular gas profile can then be
deduced from CO observations with a metallicity dependent conversion
factor recently determined (see Boselli et al., 2002 and Sec. 2.3 below).
For this reason, these galaxies do not form a complete sample
in any sense but are the ones for which the metallicity-dependent
conversion factor can be used. 
An accurate estimate
of the radial profile of atomic gas and of the star
formation rate (here deduced from \Ha \ emission line 
intensities) is also required for our purpose.
Some of the proposed instability
criteria in galactic disks involve the rotation frequency (requiring  
knowledge of the rotation curve), the stellar surface density
profile  (obtained from the H and K band profiles), and also  the
velocity dispersion of the stellar component (derived in Sec. 3.1).

The final list of galaxies, only selected to have abundance gradients
published and the other data needed for our analysis, includes 16
galaxies, 7 of them ``isolated'' and 9 belonging to the Virgo cluster
(most of the data concerning the Virgo galaxies are available via the
goldmine database, Gavazzi et. al, 2003). Note that some of the
``isolated'' galaxies are in fact part of small groups (NGC2403, 3031,
4258, 5194).

In order to estimate the degree of perturbation induced by the
interaction of a galaxy with its environment, we use the {\it HI
deficiency parameter}, defined as HI def. = log($<$HI$>$/HI), the
ratio of the average HI mass in isolated objects of similar
morphological type and linear size to the HI amount of an individual
galaxy (\citet{haynes84}). The sample includes both gas deficient (HI
def. $>$0.3) and gas rich galaxies to study whether their SFR profiles
are affected by the amount of available gas. It has been selected to
span the largest possible range in gas column density within late-type
galaxies of similar type.

\begin{table*}
{\scriptsize
	\caption[]{The sample of galaxies used in this study. Properties are given in Cols 2-6 and 
         References for the data in Cols 7-11.
         Description of the Table entries is made in Sect. 2. 
	The first half of the Table concerns nearby galaxies, 
            while the second half refers to Virgo galaxies.}
         \label{tablegeneral}
	\begin{tabular}{l c c c c c c c c c c r }
	\hline
	\hline
Galaxy      &  m$_B$ (mag) & Dist.(Mpc) & Type & log($W_C$) & HI def. & HI & CO & Z & H$_\alpha$ &  V(R) & Photometry \\
(1)         &   (2)       & (3)        & (4)  & (5)  & (6)&  (7)            &  (8)    &  (9)    &  (10)  &  (11) & (12)  \\
\hline
NGC925	    & 10.69 &  9.16 & SAB(s)d    & 2.42  &  0.0 & a  & f   & i,j   &o & p   & B,V,H\\     
NGC2403     & 8.93  &  3.22 & SAB(s)cd   & 2.47  &  0.1 & a  & h   & i,j,k &o & q,r & B,V,H,K\\     
NGC2541     & 12.26 & 11.22 & SA(s)cd    & 2.33  &  0.7 & b  & g   & j     &o & s   & U,B,V,H\\     
NGC2903     & 9.68  &  8.90 & SB(s)d     & 2.65  &  0.4 & a  & h   & i,j   &o & q,r & U,B,V,H\\     
NGC3031     & 7.89  &  3.63 & SA(s)ab    & 2.67  &  0.1 & c  & f   & j     &o & r,t & B,V,H\\       
NGC4258     & 9.10  &  7.98 & SAB(s)bc   &  2.68 &  0.5 & a  & h   & j     &o & r,u & B,V,H\\       
NGC5194     & 8.96  &  8.40 & SA(s)bc-pe & 2.76  &  0.6 & d  & h   & j     &o & r   & B,V, H\\     
\hline
NGC4254     & 10.44 & 17.00 & SA(s)c     & 2.76  &  0.0 & e  & h   & l     &o & v,w,x & B,V,H\\     
NGC4303     & 10.18 & 17.00 & SAB(rs)bc  & 2.47  & -0.1 & e  & h   & l     &n & r,x,y,z     & B,V,H\\     
NGC4321     & 10.05 & 17.00 & SAB(s)bc   & 2.73  &  0.5 & e  & h   & l     &o & r,x,y   & U,B,V,H\\     
NGC4501     & 10.36 & 17.00 & SA(rs)b    & 2.79  &  0.5 & e  & h   & l     &o & x,y,z   & U,B,V,H\\     
NGC4571     & 11.82 & 17.00 & SA(r)d     & 2.66  &  0.5 & e  & h   & l     &m & y      & B,H\\       
NGC4651     & 11.39 & 17.00 & SA(rs)c    & 2.72  & -0.2 & e  & h   & l     &m & y,z      & H\\       
NGC4654     & 11.10 & 17.00 & SAB(rs)cd  & 2.56  & -0.3 & e  & h   & l     &n & v,x,y      & B,V,H,K\\     
NGC4689     & 11.60 & 17.00 & SA(rs)bc   & 2.47  &  0.9 & e  & h   & l     &n & v,x,y      & B,V,K\\     
NGC4713     & 12.19 & 17.00 & SAB(rs)d   & 2.35  & -0.4 & e  & --  & l     &m & v,y     & V,H\\      
\hline
\end {tabular}

{\it References}: 
\citet{wevers86} [a], \citet{broeils94} [b], \citet{rots75} [c],  
\citet{rand92} [d],  \citet{warmels86} [e],
\citet{sage93} [f], \citet{braine93} [g], \citet{young95} [h],
\citet{vanzee98} [i], \citet{zaritsky94} [j], \citet{garnett97} [k], \citet{skillman96} [l]
\citet{martin01} [m], Hippelein et al., in preparation [n], Boselli \& Gavazzi (2002) [o],
\citet{pisano98} [p], \citet{begeman87} [q], \citet{sofue99} [r], \citet{vega01} [s], 
\citet{adler96}[t], \citet{vanderkruit74} [u], \citet{sperandio95} [v], \citet{chincarini85} [w].
\citet{guha88} [x], \citet{rubin99} [y], \citet{distefano90} [z].
The images from which the photometric data are derived are presented for the 
Virgo galaxies in Boselli et al. (1997, 2000) (H), and Boselli et al. (2003) (optical). For the
other galaxies in the infrared: Pierini et al. (1997), Boselli et al. (2000), Zibetti et al., in 
preparation. The optical data will be presented in a following paper making a larger use of them
(Boissier et al., in preparation).
}
\end{table*}

Information on the galaxies of the adopted sample is given in
Table \ref{tablegeneral}.  Column 1 gives the name of the galaxy and
Column 2 the blue magnitude. 
Column 3 gives the galaxy distance. A distance of 17
Mpc is adopted for all the Virgo galaxies. 
We adopt the distance
deduced from the brightest stars of NGC2903 \citep{drozdovsky00}
and the planetary nebulae distance of NGC5194 by
\citet{feldmeier97}. For the other galaxies of the sample, Cepheid distances were
deduced in the framework of  the HST key project \citep{freedman01}.
Column 4 gives the morphological type.
Column 5 gives the logarithm of the HI
velocity width $W_C$ (in km s$^{-1}$), corrected for inclination as in
Gavazzi (1987).
Column 6 gives the HI-deficiency,    
computed as in \citet{haynes84}, from the 
integrated HI masses \citep{huchtmeier89} and 
the blue major axes as given in the Uppsala General Catalogue
\citep{nilson73}.

For each of our galaxies, Table 1 provides also references for the adopted
profiles of neutral gas, 
CO emissivity and oxygen abundance data (columns 7, 8, and 9, respectively).
Column 10 presents  the sources of the H$\alpha$ data.  
For the majority of the sample we adopt the results of Boselli \& Gavazzi (2002);
for three galaxies  (m) we use the data kindly made available by 
\cite{martin01} and for three others
(n) data obtained by Hippelein et al. (in preparation).
\label{secobsdesc}

Column 11 gives references for rotation curves.  Finally, Column 12
gives the photometric bands for which we have images, allowing to
derive corresponding stellar surface density profiles.

For NGC3031 and NGC4258, we used the H band profile provided
in the 2mass large galaxies atlas: 
http://www.ipac.caltech.edu/2mass/gallery/largegal/
\citep{jarrett02}.

\subsection{Atomic gas profiles}

\label{sechi}

With the exception of NGC5194 (HI-map obtained with the VLA), the data
were obtained with the Westerbork Radio Telescope (see
caption of Table \ref{tablegeneral}). The HI
profile given by \citet{broeils94} was obtained along the major axis
of the galaxy. The data by \citet{wevers86}, \citet{rots75} and 
\citet{rand92} were obtained by integrating HI maps over concentric ellipses.

The data concerning the Virgo galaxies are taken from the thesis of
Warmels \citep{warmels86}. His data are two dimensional maps for 
NGC4321, NGC4501, and NGC4651. For the other galaxies, the data were obtained
along one (and sometimes two) resolution axis. The maps were integrated onto
one resolution axis, and the same procedure was applied to the whole sample
to obtain the HI density profiles from the data along one axis.

In all the cases, the HI profiles correspond to a ``face-on'' orientation.
A typical uncertainty of 20 \% should be quoted for all the HI profiles,
which is  smaller than the uncertainties of H$_2$ profiles
(see next section).

\subsection{Molecular gas profiles}

\label{secco2h2}

The adopted CO intensity profiles 
have been  corrected for inclination. 
The spatial resolution of the
CO observations is relatively poor (45'') and
the measurements were made only along the
major axis of each galaxy. 
On the other hand, the relatively large beams average the CO
emission on larger scales and are less affected by small-scale clumpiness.

In their recent study concerning the star formation rate in galactic disks, 
Wong \& Blitz (2002) used high resolution
profiles obtained with the BIMA interferometer
\citep{regan01} for 7 disk galaxies. 
The observations of \citet{regan01} concern 15 galaxies from 
which four are in common with our sample (NGC 2903, 4258, 5194, 
4321). Their profiles (Figure 3 in Regan et al. 2001)
differ from ours in the innermost parts of the galaxies, but we are only
interested in galactic discs and not in bulges in this work, 
so that this difference is not of importance for our conclusions.

The low resolution CO profiles adopted in this work are obviously
missing any small scale structure seen in the high resolution profiles
of Regan et al. (2001). Our profiles, 
however, agree well with the low-resolution one also shown in Regan et
al. (2001; Fig. 3), which represent a good estimate of the average CO
profile in most cases.
Among  the four galaxies that are common in the two samples,
only NGC4258 presents really important variations ($\sim$ one magnitude)
on small scales. 

\begin{figure}
\centering
\includegraphics[width=0.5\textwidth]{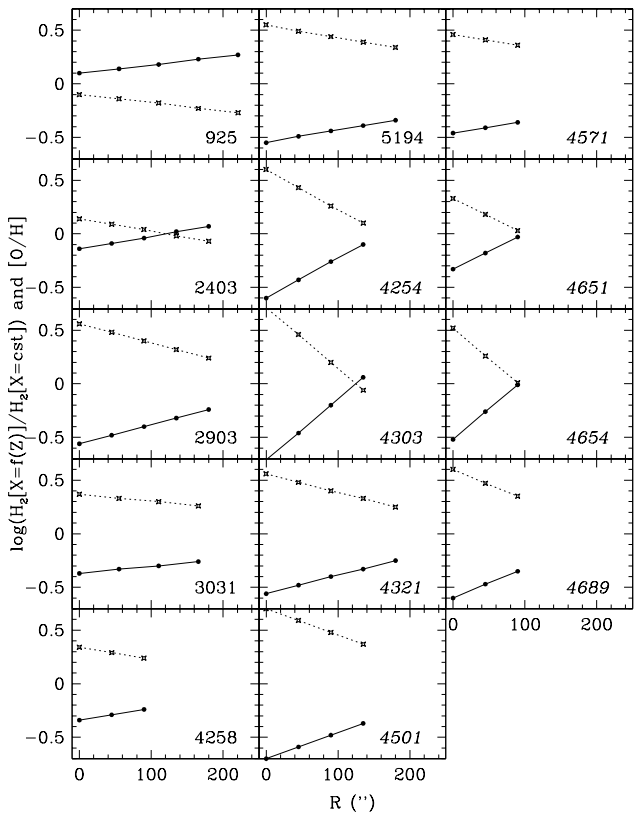} 
\caption{\label{figCOfact} 
{\it Solid curves:} Amount of molecular gas obtained by 
our method (i.e. with metallicity dependent 
conversion factor of CO to H$_2$), divided by the amount of molecular gas 
obtained with a constant conversion factor. Results are plotted as a function
of galactocentric radius. {\it Dotted curves:}
The corresponding metallicity profiles [O/H]=log(O/H)-log(O/H)$_{\odot}$. 
As in the following figures, the name of the Virgo galaxies is 
indicated in an italic font.
} 
\end{figure}

The uncertainty  on the H$_2$ content is much larger than the one of the CO  data; it is
also more difficult to quantify, since it depends on the  poorly
known factor  X converting CO to H$_2$. In a recent work \citet{boselli02b},
based on spectro-photometric data available for a small number 
of nearby galaxies and on a larger sample of late type galaxies with
multi-frequency data,
found that  X depends on metallicity (or luminosity) 
and they parameterised that dependence as:
\begin{equation}
\log X \ = \ -1.01 \ (12. + \log (O/H)) \ + \ 29.28
\label{eqX}
\end{equation}
i.e. the conversion factor is essentially inversely 
proportional to metallicity.
While this correlation was found for integrated values over whole galaxies, 
we make the assumption that it is also valid for radial profiles. 
This assumption agrees with the analysis of M51 done by Nakai \& 
Kuno (1995) that, based on the measured radial varaition of the 
dust to gas ratio, brought to similar results.

We transformed CO data into molecular 
H$_2$ surface density using the metallicity dependent conversion 
factor of Equ. 1.
We note that the resulting H$_2$ profile is much poorer in molecular
gas, especially in inner galactic regions, than profiles obtained
with a constant factor. In Fig. 1 we show the ratio of the
two profiles i.e. the one obtained with a metallicity dependent X divided by the
one obtained with a constant X.
It is clearly seen that differences by a factor of 2-3 in the  
inner galaxy are usually obtained.

We note that \citet{wong02} use a constant conversion factor X, as in fact 
most studies of SFR in the literature.  This explains part of the
differences between our study and others (see discussion in
Sec. \ref{seclawfits}).

\begin{figure*}[t]
\centering
\includegraphics[angle=-90,width=\textwidth]{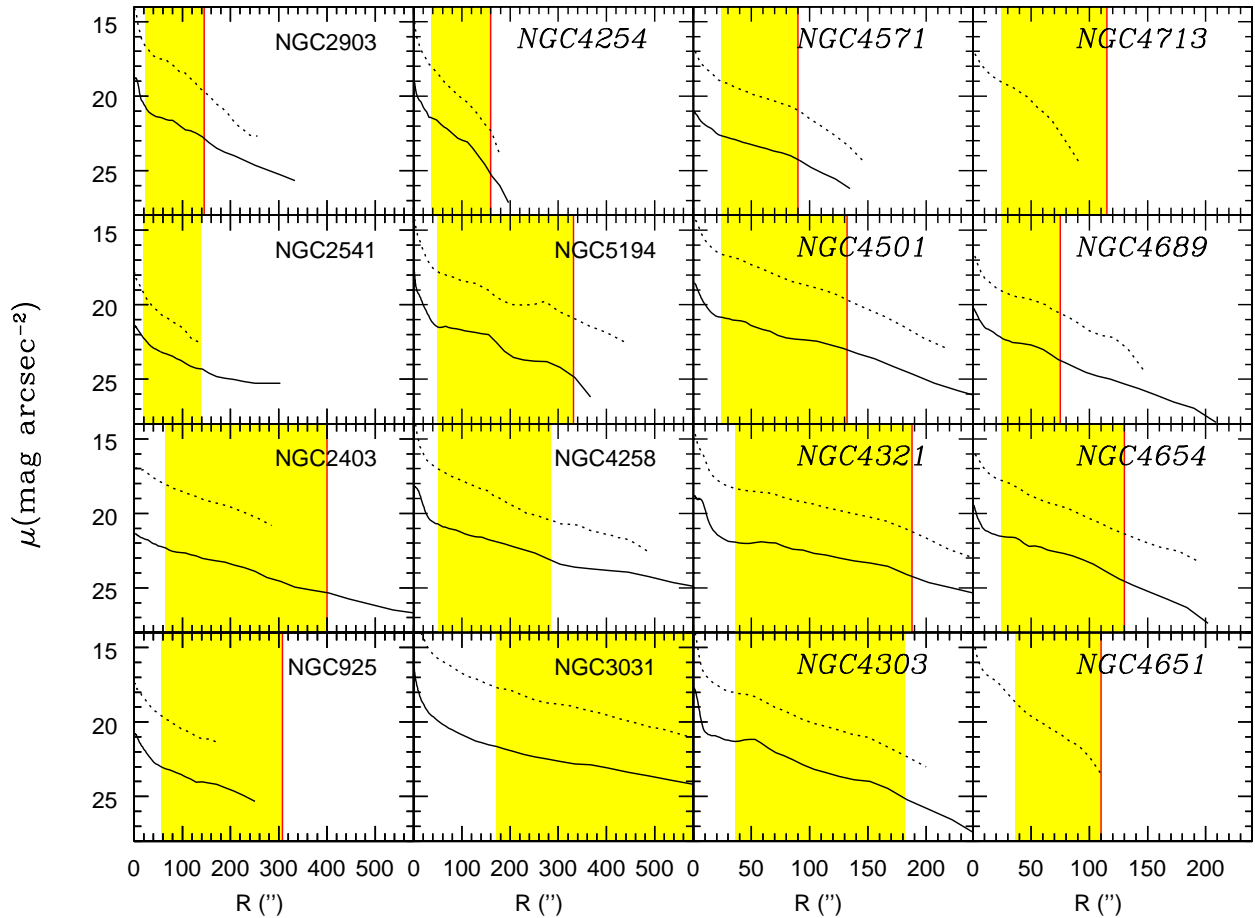} 
\caption{\label{figprofphot} 
Surface brightness radial profiles in the B-band ({\it solid curves}) and in the H-band 
({\it dotted curves})
for the galaxies of our sample. The {\it grey area}  is the ``disk'' region, used for the
study of the SFR properties. Its outer radius is either the \citet{martin01} ``cut-off'' (vertical
line, as in Fig. 5) or our own ``cut-off'' (Sec. 2.8). Its inner radius is the point where
the bulge starts dominating over the disk (Sec. 2.6).
}
\end{figure*}

\subsection{Star Formation Rate profiles}

\label{secsfrprof}

The star formation rate (SFR) profiles were obtained by
subtracting the radial profiles of the continuum near \Ha{}
from the \Ha+[NII] radial profiles. For three of our galaxies,
we used \Ha{} profiles of \citet{martin01}, kindly provided
by the authors.  

The \Ha{} flux was integrated along elliptical annuli with the 
same PA and inclination as for the HI (see Table \ref{tablepa}
for the adopted values). It was corrected for inclination, 
and translated to a star formation rate as in \citet{boselli01}. 
\Ha+[NII] fluxes were corrected for [NII] contamination
and internal extinction (see Boselli et al. 2001).
The conversion from {\it extinction corrected} 
\Ha{} flux to star formation rate is obtained
assuming a power law  IMF of slope -2.5 between 0.1 and 80 \Msol:
$$ SFR ({\rm M_{\odot}/yr}) \ = \ 0.86 \ 10^{-41} \ L_{H\alpha}({\rm erg/s}) $$

As estimated in Boselli et al. (2001),
the uncertainties in the IMF, extinction and [NII] contamination correction
result in an overall uncertainty of the absolute SFR of a factor $\sim$ 3. 
\begin{table}
{\scriptsize
	\caption[]{Parameters adopted for the determination of the \Ha \ and photometric profiles}
         \label{tablepa}
	\begin{tabular}{l c c }
	\hline
	\hline
Galaxy      &  Inclination($^o$) & PA($^o$) \\
\hline
NGC925	    &  54.00 &  102  \\     
NGC2403     &  60.00 &  125  \\
NGC2541     &  66.80 &  -29  \\
NGC2903     &  60.00 &  29   \\
NGC3031     &  68.00 &  332  \\
NGC4258     &  63.00 &  330  \\
NGC5194     &  20.00 &  163  \\
\hline
NGC4254   &  24.51 &  56   \\
NGC4303   &  35.99 &  7    \\
NGC4321   &  27.37 &  153  \\
NGC4501   &  58.62 &  140  \\
NGC4571   &  21.09 &  55   \\
NGC4651   &  45.24 &  71   \\
NGC4654   &  59.05 &  128  \\
NGC4689   &  41.01 &  55   \\
NGC4713   &  47.80 &  100  \\
\hline
	\end {tabular}
}
\end{table}

We note that
\citet{wong02}  applied radially dependent extinction corrections to \Ha{}, using the
observed gas profiles, an assumed dust to gas ratio, and a given geometry
between HI, H$_2$ and the stars. We prefer here to apply a simple correction
for the whole galaxy since, as stressed in \citet{martin01}, 
``radial gradients in internal extinction are typically not strong compared to
the scatter in extinction at a given radius''. Moreover, the approach
of \citet{wong02} does not take into account the existence of an abundance
gradient within the discs, which is likely to affect the dust to gas ratio
in different ways at different radii.

\subsection {Metallicity profiles}

In order to compute the conversion factor X from CO to H$_2$, we need the 
metallicity as a function of radius.\citet{zaritsky94},
\citet{vanzee98},  \citet{garnett97} and \citet{skillman96}
measured with the same method the oxygen abundances in HII regions
located at various galacto-centric radii in the sample of our
galaxies.  From their data, they derived an abundance gradient (slope
and intersect at the centre of the galaxy) that we use to estimate the
metallicity at each radius (see Fig. \ref{figCOfact}).

\subsection{ Photometry}

The photometric profiles  obtained in the framework
of our project were treated in a similar way as the \Ha{}
profiles (i.e. they were integrated over ellipses with the same position angle and
inclination, given in table 2). In Fig.  \ref{figprofphot}, we present the B and H-band
profiles, which are the most useful for the present work. 

We derive a scalelength $h_B$ by
an exponential fit to the disc part of the B-band profiles 
(to be used in section \ref{secbottema}). The H-band
profiles, unaffected by dust extinction, have been used 
in order to
estimate the stellar mass density, assuming the same mass to light
ratio as in \citet{bo2pg}. In the case of NGC 4689 the
stellar profile was scaled from the K band profile (since the
H-band profile was not available), taking into account the total magnitudes
in H and K bands, respectively H=8.32 and K=7.76 (Boselli et al., 1997).

\begin{figure*}
\centering
\includegraphics[angle=-90,width=\textwidth]{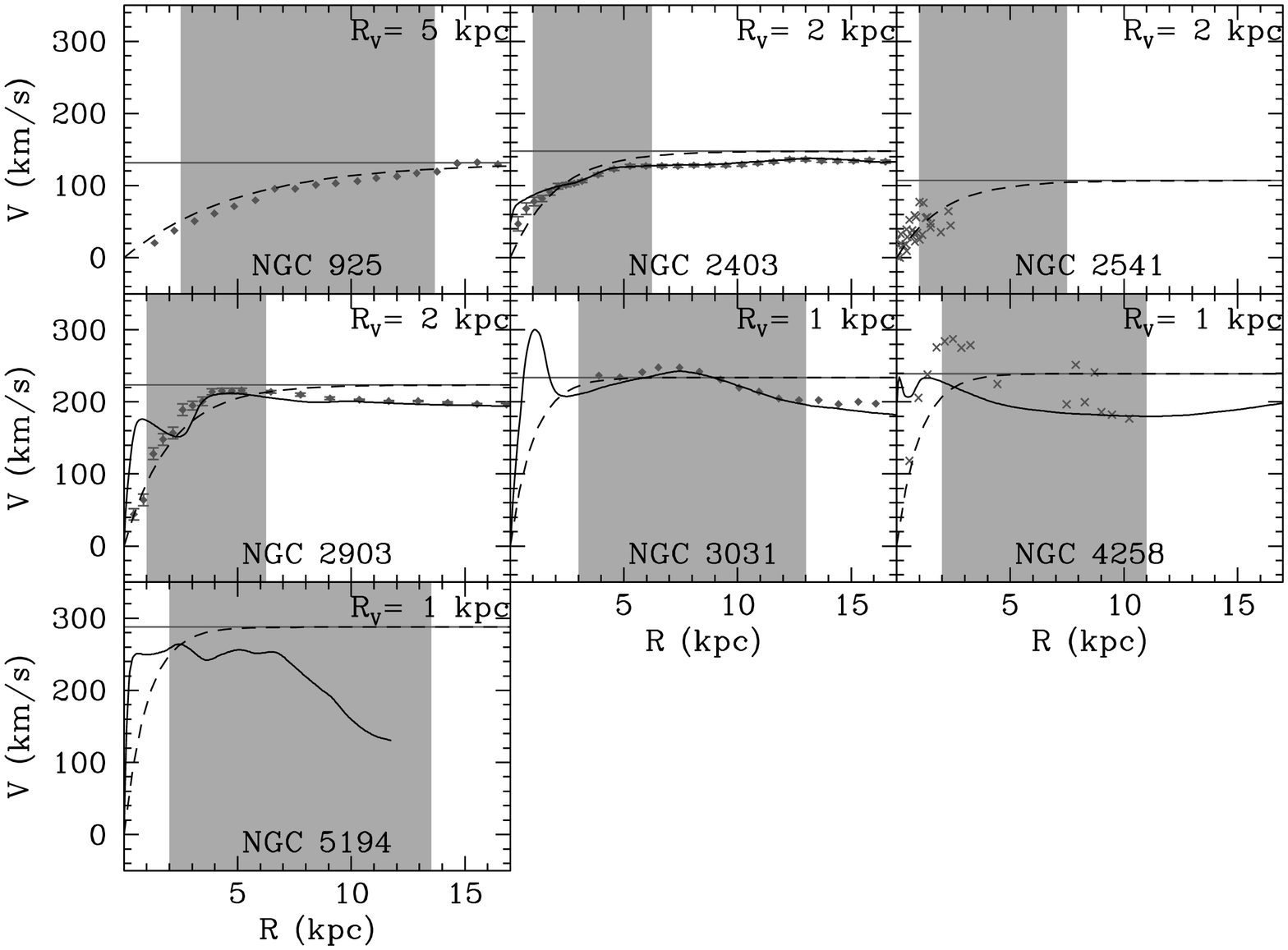} 
\caption{\label{rotanonvirgo} 
Rotation curves for the isolated galaxies of our sample.
{\it Diamonds}: HI data (\citet{pisano98}
for NGC925, \citet{begeman87} for NGC2403 and NGC2903, 
\citet{adler96} for NGC3031).
{\it Crosses} : Optical slit spectra 
(\citet{vanderkruit74} for NGC4258, \citet{vega01}  for NGC2541.
The {\it solid curve} indicates the composite rotation curve of \citet{sofue99}.
The {\it horizontal line} indicates half the value of the HI line width.
The {\it dashed curve} is the function of Equ. \ref{pseudorota},
with the corresponding value of $R_V$ indicated for each galaxy.
{\it Shaded areas} indicate the disk region, used for the study of the star formation.
} 
\end{figure*}

\begin{figure*}
\centering
\includegraphics[angle=-90,width=\textwidth]{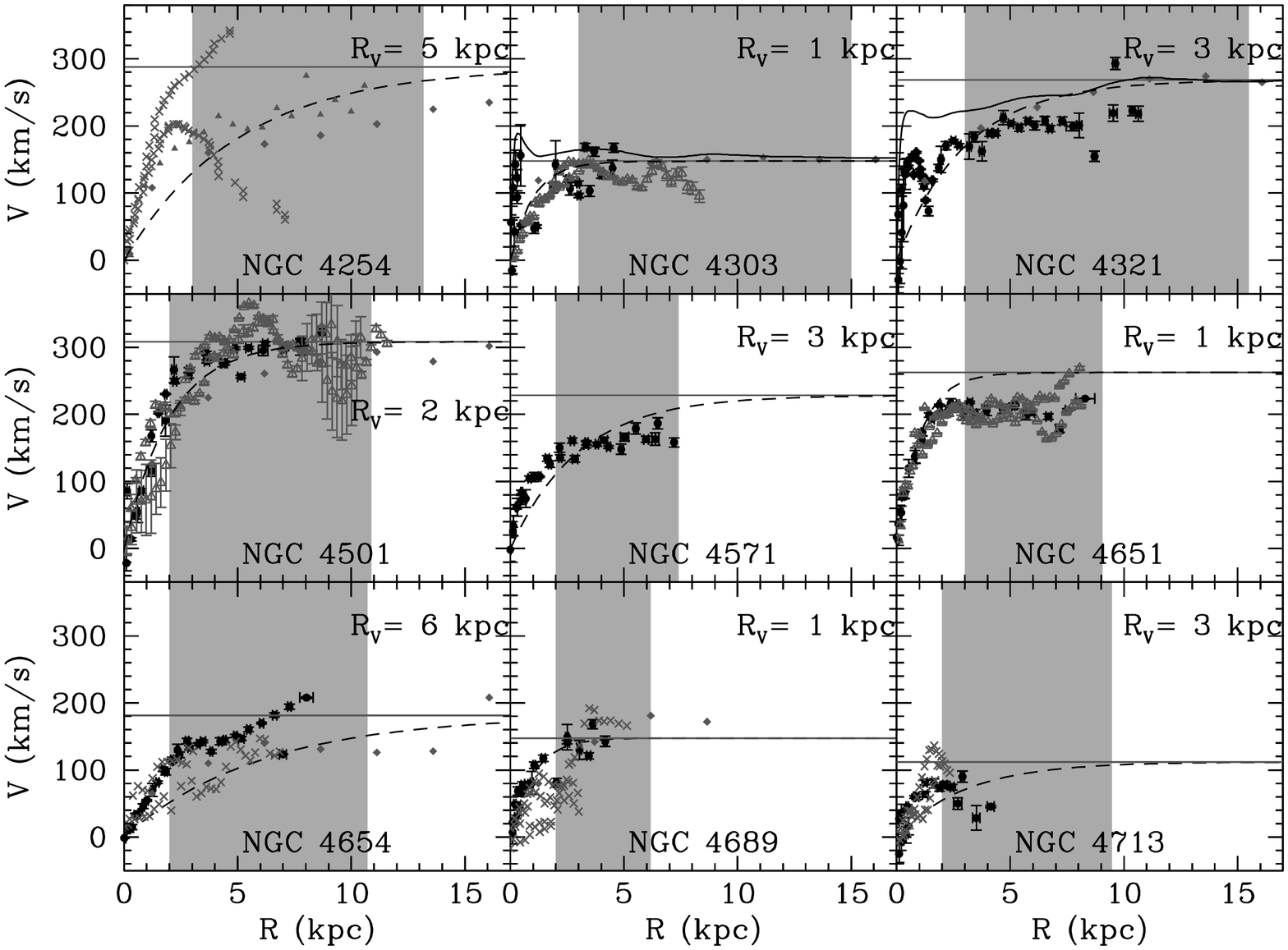} 
\caption{\label{rotavirgo} 
Rotation curves for the Virgo galaxies of our sample.
{\it Diamonds}: HI data \citep{guha88}.
Data of optical slit spectra  are from \citet{rubin99} ({\it filled circles})
and \citet{sperandio95} ({\it open circles)}.
The {\it solid curve} indicates the composite rotation curve of \citet{sofue99}.
As in Fig. 3,
the {\it horizontal line} indicates half the value of the HI line width,
the {\it dashed curve} is the function of Equ. \ref{pseudorota},
with the corresponding value of $R_V$ indicated for each galaxy,
{\it shaded areas} indicate the disk region, used for the study of the star formation.
} 
\end{figure*}

An inspection of the profiles (Fig. \ref{figprofphot}) shows
that in the inner 1 or 2 kpc they present departures from the
exponential disc profile, obviously due to the presence of a bulge. 
These inner regions are ignored in our analysis.

\subsection{Rotation curves}
\label{secrotacurve}

The rotation curves of the galaxies
of our sample are obtained through a variety of sources: HI rotation curves 
of \citet{pisano98}, \citet{begeman87},
\citet{adler96}, Guhathakurta et al. (1988),
rotation curves derived from optical spectroscopy by
\citet{vanderkruit74},
\citet{vega01},  \citet{chincarini85}, \citet{distefano90} and for the
Virgo galaxies by \citet{rubin99} and Sperandio et al. (1995).
For seven of our  galaxies, high resolution curves are given by \citet{sofue99},
resulting from a combination of CO, optical and HI data.

The adopted rotation curves  are presented in figures \ref{rotanonvirgo} and
\ref{rotavirgo} for isolated and Virgo galaxies, respectively. In those
figures 
we also indicate by a horizontal line half the value of $W_C$ (the HI line width),
in order to show the difference with a flat rotation curve. This constant value usually
represents well the ``plateau'' of the rotation curves, which is not 
reached at the same radius for all the galaxies.  It should be noted that
the data originate from
different sources (i.e. studies with different angular resolutions) and 
present a substantial amount of scatter. For the purpose of homogeneity
we approximated the rotation curves with the simple function:
\begin{equation}
\label{pseudorota}
V(R) \ = \ 0.5 \ W_C \ (1-exp^{-R/R_V})
\end{equation}
At large radii, this approximation produces a plateau at the value
$W_C$ given by the HI line width. At small  radii, the  parameter $R_V$
controls the rise of curve.  $R_V$ was adjusted for 
each galaxy, allowing the curve of equation \ref{pseudorota} 
to reproduce satisfactorily the observed
one. This simple representation has the advantages that
i) it provides a uniform description of the rotation curve for all galaxies, 
ii) it has only one
free parameter and iii) it can be derived analytically, a  useful property for
the study of stability  criteria in disks (see section \ref{secthres}).

\begin{figure*}
\centering
\includegraphics[angle=-90,width=\textwidth]{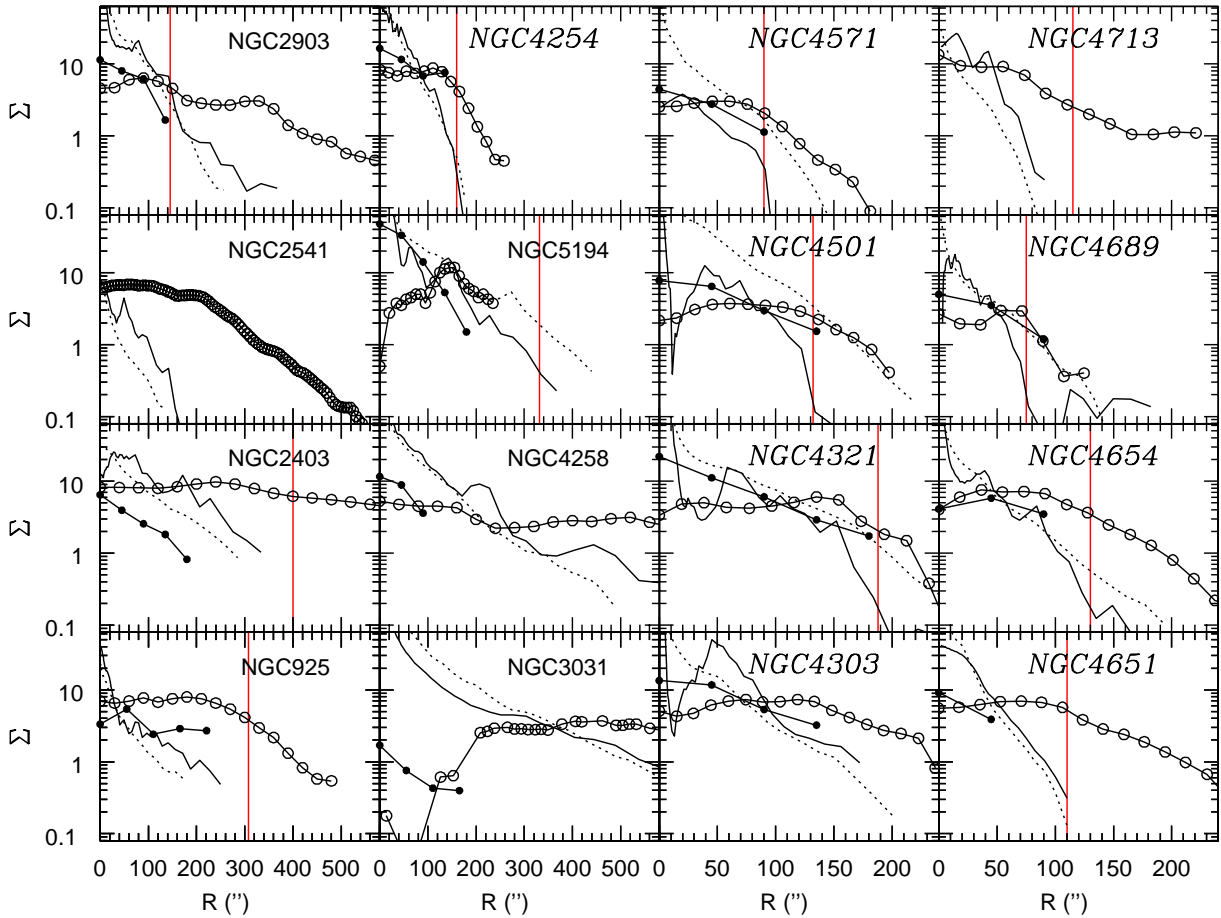} 
\caption{\label{figprofobs} 
Surface density radial profiles of HI ({\it open circles}), H$_2$ ({\it filled circles}),
Star Formation Rate ({\it solid curves}), and Stellar density ({\it dotted curves}). 
The surface densities of HI and H$_2$ are in
\Msol{} pc$^{-2}$, of stars in 10 \Msol{} pc$^{-2}$ and of the 
star formation rate in \Msol{} pc$^{-2}$ Gyr$^{-1}$.
The {\it vertical line} in some of the panels indicates the SFR ``cut-off' found by \citet{martin01}.
In most cases that line coincides with (or is very close to) a region of
steep decline in our own SFR profile.
} 
\end{figure*}

\subsection{Comments}

Our H$_2$, HI and SFR surface density profiles are displayed in Fig. 5. 
Several points should be noted concerning these profiles:

- Resolution is quite different in the profiles of the
three major quantities studied. It is low in the case of H$_2$, where only 3-5 points are
available along the disk; it is satisfactory in the case of HI, with more than 20
points per galaxy, on average ; and it is excellent in the case of
the \Ha{} profile, where a large number of features appear on the corresponding curves,
resulting from azimuthally averaging across the spiral arms.
Even if we made an enormous effort of homogenisation, the data
are not as deep at all wavelength, in particular for H$_2$.

- In most cases, the molecular gas dominates the inner disk, with the exception of NGC2403,
NGC4654 and NGC925 (three HI-deficient disks). However, its surface density rarely exceeds
10 \Msol/pc$^2$. Only in one case (NGC5194) does it 
become as large as 40 \Msol/pc$^2$.

- In general, the HI disk is more extended than the SFR profile and the 
latter is more extended than the H$_2$ profile.
In more than half of the cases 
the SFR becomes negligible (or declines
strongly) at the outer limit of the molecular disk. 
The accurate determination of the molecular gas profile, however,
is strongly limited by the difficulty of detecting CO in the 
low-metallicity, cold outer disc.
We shall return to this
point concerning a ``cut-off'' of the SFR in the next section.

\section{Stability criteria and SFR threshold}

\label{secthres}
\subsection{ Theoretical background}

The data presented in Figs. 2, 3, 4 and 5 allow 
the question of a threshold of
the SFR in disk galaxies to be assessed. 
Such a threshold has been
 suggested on both observational and theoretical
grounds. On the observational side, support comes from the fact that the gaseous layer
(HI) is often observed  to extend much beyond the stellar disk. On the other hand, the
formulation of local instability criteria for differentially rotating 
disks (Toomre 1964) leads naturally to the idea that large scale star formation 
may  occur only above a critical gas surface density. According to Quirk (1972) 
this happens when the Toomre parameter
\begin{equation}
Q = {{\Sigma_{CRIT}}\over{\Sigma_{GAS}}} \ < \ 1 
\label{eqtoomre}
\end{equation}
where $\Sigma_{GAS}$ is 
the  gas surface density and $\Sigma_{CRIT}$ is given by
\begin{equation}
\Sigma_{CRIT} = 
{{\kappa \ \sigma_{GAS}}\over{\pi G}} 
\label{eqcrit}
\end{equation}
where $\sigma_{GAS}$ is the local gas velocity dispersion and
$\kappa$ is the local epicyclic frequency, given by
\begin{equation}
\kappa^2 = {{2 V_C}\over{R}} \ ({{dV_C}\over{dR}} + {{V_C}\over{R}}) 
\label{eqkappa}
\end{equation}
where $V_C$ is the gas circular velocity of the galaxy at radius $R$.  For flat rotation curves
one has $\kappa$=$\sqrt{2} \ V_C/R \ = \ \sqrt{2} \ \Omega$, where $\Omega$ is the
rotational frequency of the disk.

The gas velocity dispersion in the radial direction is of the order of
5-10 km s$^{-1}$, as derived by observations of e.g. \citet{sanders84} in the
Milky Way.  In external galaxies, gas velocity dispersions obtained by
\citet{dickey90} and \citet{boulanger92} span a similar range of
values and they present rather flat profiles, i.e. independent of
radius (except in the inner parts of the disks where they increase,
but by a factor never exceeding $\sim $2). On theoretical grounds the
gas velocity dispersion is expected to vary little in self-regulated
disks (Silk 1997).  In that case, one obtains for disks with flat
rotation curves : $\Sigma_{CRIT} \propto \kappa
\propto R^{-1}$. Since this gradient of $\Sigma_{CRIT}$ is shallower than the observed 
decline of the gas surface density, it is expected that $Q>1$ at some ``cut-off'' radius,
resulting naturally in a threshold of the SFR.

These ideas were studied in a seminal paper by Kennicutt (1989), with
a sample of 15 (mostly late type) spirals. By using a constant gas
velocity dispersion $\sigma_{GAS}$=6 km s$^{-1}$ he found that star
formation declines rapidly in disk regions with $Q>$1.6, i.e. for
$\Sigma_{GAS}<$0.6 $\Sigma_{CRIT}$. He also found that $Q$ varies
little within the star forming disk, an indication of self-regulating
star formation.

\citet{wang94} proposed to take into account the stellar contribution to the
instability by rewriting the critical density as:
\begin{equation}
\Sigma_{CRIT} = \gamma {{\kappa \sigma_{GAS}}\over{\pi G}}, {\rm with} \ 
\gamma=\left(1+\frac{\Sigma_* \sigma_{GAS}}{\Sigma_{GAS} \sigma_*}\right)^{-1} 
\label{eqcritaeff}
\end{equation}
where $\Sigma_*$ and $\sigma_*$ are, respectively, the stellar surface
density and the {\it radial} stellar velocity dispersion.

In the rest of Sec. 3 we study the instability criterion of 
equation \ref{eqtoomre} for galactic disks by
using both definitions (\ref{eqcrit}) and (\ref{eqcritaeff}) for $\Sigma_{CRIT}$. 
In the former case, we adopt
a constant value $\sigma_{GAS}$= 6 km s$^{-1}$ for the gas velocity dispersion. In the latter, a
knowledge of the radial stellar velocity dispersion is also required.

As shown by \citet{bottema93}, the radial stellar velocity
dispersion of a disk 
is exponentially decreasing with the galactocentric radius; the corresponding
scalelength is equal to twice the one observed in the B- band:
\begin{equation}
\sigma_*(R)=\sigma_0 \exp(-R/2h_B) \label{eqsigr}
\end{equation}

\begin{figure}
\centering
\includegraphics[angle=-90,width=0.5\textwidth]{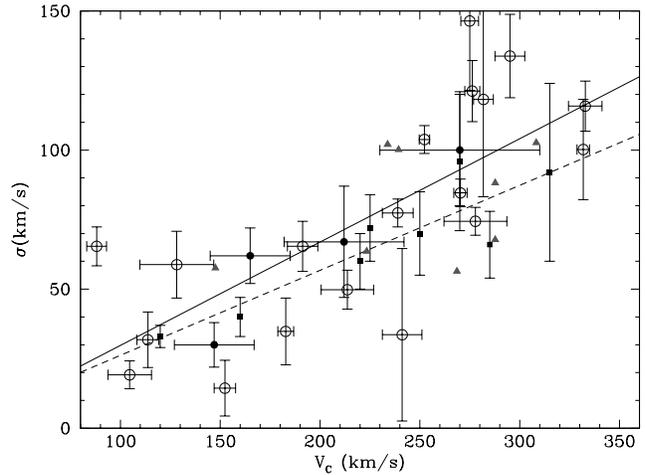} 
\caption{\label{figbottema} 
The radial stellar velocity dispersion at one scalelength for the inclined
discs (filled squares) and the vertical one at R=0 for the face-on discs
(filled circles) of \citet{bottema93}. Open circles are the recent 
data of \citet{vega01}. The dashed and solid lines show respectively
a simple fit to the  data of \citet{bottema93} and \citet{vega01}.
The triangles correspond to galaxies of our sample for which a velocity
dispersion is given in \citet{mcelroy95}
}
\end{figure}

Moreover, \citet{bottema93} showed that the value of the radial stellar velocity dispersion
at one scalelength is comparable to the vertical stellar velocity dispersion at the
centre of the galaxy.
He also showed that the stellar velocity dispersion in a galactic disk
is correlated with its rotational velocity $V_C$ (Fig. \ref{figbottema}).
We computed a simple fit to his data (for the more reliable inclined galaxies) to
obtain:
\begin{equation}
\sigma_*(R=h_B) {\rm [km s^{-1}]} \ = \ -4.3 \ + \ 0.3 \ Vc \ {\rm [km s^{-1}]} \label{eqsigv}
\end{equation}
i.e. the velocity dispersion of the stellar component is $\sim$1/3 of the
rotational velocity of the disk.
In the same figure we display  the recent data obtained by \citet{vega01} 
concerning the kinematics of 20 disc galaxies. Despite some scatter, these
data confirm the trend found by \citet{bottema93}.
For 8 galaxies of our sample, central dispersion velocities are given
by \citet{mcelroy95}. Assuming the same radial dependence as in Equ. \ref{eqsigr},
we computed the corresponding dispersion at one scalelength and plotted it also
in Fig. \ref{figbottema}; the results for the disks of our sample
are clearly in agreement with
the aforementioned  trend. We are using then in the following these stellar
velocity dispersions to calculate the value of $\Sigma_{CRIT}$, according to Equ.
\ref{eqcritaeff}

\label{secbottema}

\subsection{Application to the Milky Way}

\begin{figure}
\centering
\includegraphics[width=0.5\textwidth]{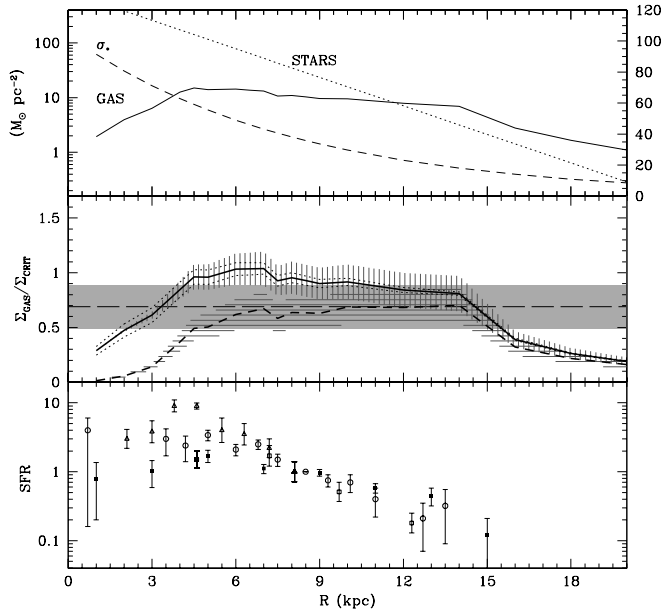} 
\caption{\label{fib} \label{figcritmw}
Gas, stars and star formation in the Milky Way disk.
\underline{Upper panel}:
observed profiles of the total gas surface density ({\it solid curve},
H$_2$+HI, increased by 40\% to include He contribution, the H$_2$
being derived from CO with a metallicity-dependent conversion factor),
the stellar profile ({\it dotted curve} exponential, with scalelength
of 2.5 kpc and normalised to 42 M$_{\odot}$/pc$^2$ in the solar
neighborhood), and the stellar radial velocity dispersion ({\it dashed
curve}, calculated according to equations \ref{eqsigr} and
\ref{eqsigv}, in km s$^{-1}$ as indicated on the right axis).
\underline{Middle panel}: profile of  the gas surface density to the 
critical density for
two cases: first ({\it dashed curve} within horizontally hatched area,
indicating the uncertainties on gas values), by using Eq. \ref{eqcrit}
for $\Sigma_{CRIT}$ (i.e. by taking into account only the gas velocity
dispersion, assumed here to be 6 km s$^{-1}$ at all radii); secondly ({\it
solid curve} within vertically hatched area, indicating the
uncertainties on gas values) by using Eq. \ref{eqcritaeff} for
$\Sigma_{CRIT}$ (i.e. including the contribution of the stellar disc
to the instability criterion). The {\it shaded area} indicates the
value 0.69$\pm$ 0.20 that Martin \& Kennicutt (2001) found for the star
formation threshold in their sample of galactic disks.
\underline{Bottom panel}: SFR profile in the Milky Way (from various tracers), normalised to 
its value in the Solar Neighborhood.
}
\end{figure}

Before studying the above stability criteria with our sample of disk galaxies,
we present in Fig. 7 an application to the Milky way disk. In the upper panel we show the
observed profiles of the total gas surface density (H$_2$+HI,
increased by 40\% to include He contribution), the stellar profile (exponential, with
scalelength of 2.5 kpc and normalised to 42 M$_{\odot}$/pc$^2$ in the solar
neighborhood), and the profile of the
stellar radial velocity dispersion, calculated according to the
prescriptions of the previous paragraph and Bottema's (1993) data, 
by combining equations \ref{eqsigr} and \ref{eqsigv}. 
In the middle panel, we show the profile of the ratio 
of the gas surface density
to the critical density, which is
calculated either by Eq. \ref{eqcrit} (i.e. by taking into account
only the gas velocity dispersion, assumed here to be 6 km s$^{-1}$ at all radii) 
or by using Eq. \ref{eqcritaeff}
(i.e. by taking into account both the gaseous and the stellar  components).

It can be seen that in the former case, the ratio
\St \ remains roughly constant at an average
value of $\sim$0.65, between 4 and 14 kpc, i.e in a large part of the
star forming disk ; this value is very close to the value of 0.69$\pm$0.20
that Martin \& Kennicutt (2001) found as defining the SFR threshold in
their sample of spiral galaxies (also shown as {\it grey shaded area} in Fig. 7). 
However, in the inner disk, i.e. in the ``molecular ring'' where
most of the star formation activity is concentrated, the values
of $\Sigma_{GAS}/\Sigma_{CRIT}$ are {\it below} the threshold found by
Martin \& Kennicutt (2001). The situation improves if 
the stellar radial dispersion profile is taken into account ({\it solid  curve} in
the middle panel of Fig. 7). Since, in the molecular ring one has
$\Sigma_*\Sigma_{GAS}\sim$10 and  $\sigma_{GAS}\sigma_*\sim$0.1,
the resulting value of $\gamma$ (Equ. 6) is $\sim$1/2 and 
$\Sigma_{GAS}/\Sigma_{CRIT}$ increases in the inner Galaxy, up to
the level of $\sim$1 in the molecular ring.

In the lower panel of Fig. 7 we present the SFR
profile of the Milky Way (normalised to the solar neighborhood SFR
value), as given by various tracers (see
Boissier \& Prantzos 1999 for references to original data). 
It can be seen that the
outer radius of the SFR profile of the Milky Way coincides with the
radius of a steep decline in the $\Sigma_{GAS}/\Sigma_{CRIT}$ profile;
this conclusion is valid for both ways of calculating $\Sigma_{CRIT}$.

To summarise, the Milky Way data 

(i) support the idea of large scale star formation according to the Toomre
instability criterion of Eq. (2), since the value of $ \Sigma_{GAS}/\Sigma_{CRIT}$
is $\sim$constant across the star forming disk,

(ii) are consistent with the Martin \& Kennicutt
(2001) data for a threshold of star formation in external spirals 
(at a value of  $ \Sigma_{GAS}/\Sigma_{CRIT}$$\sim$0.7$\pm$0.2 in most
of the star forming disk) and

(iii) favor the idea that the stellar component should also be used in the
instability criterion (at least in the inner disk).
This confirms the results of Wang and Silk (1994) who proposed
this idea and tested it in the Milky Way.

Unfortunately, the situation with the other disks of our sample is
not as clear-cut.

\begin{figure*}
\centering
\includegraphics[angle=-90,width=\textwidth]{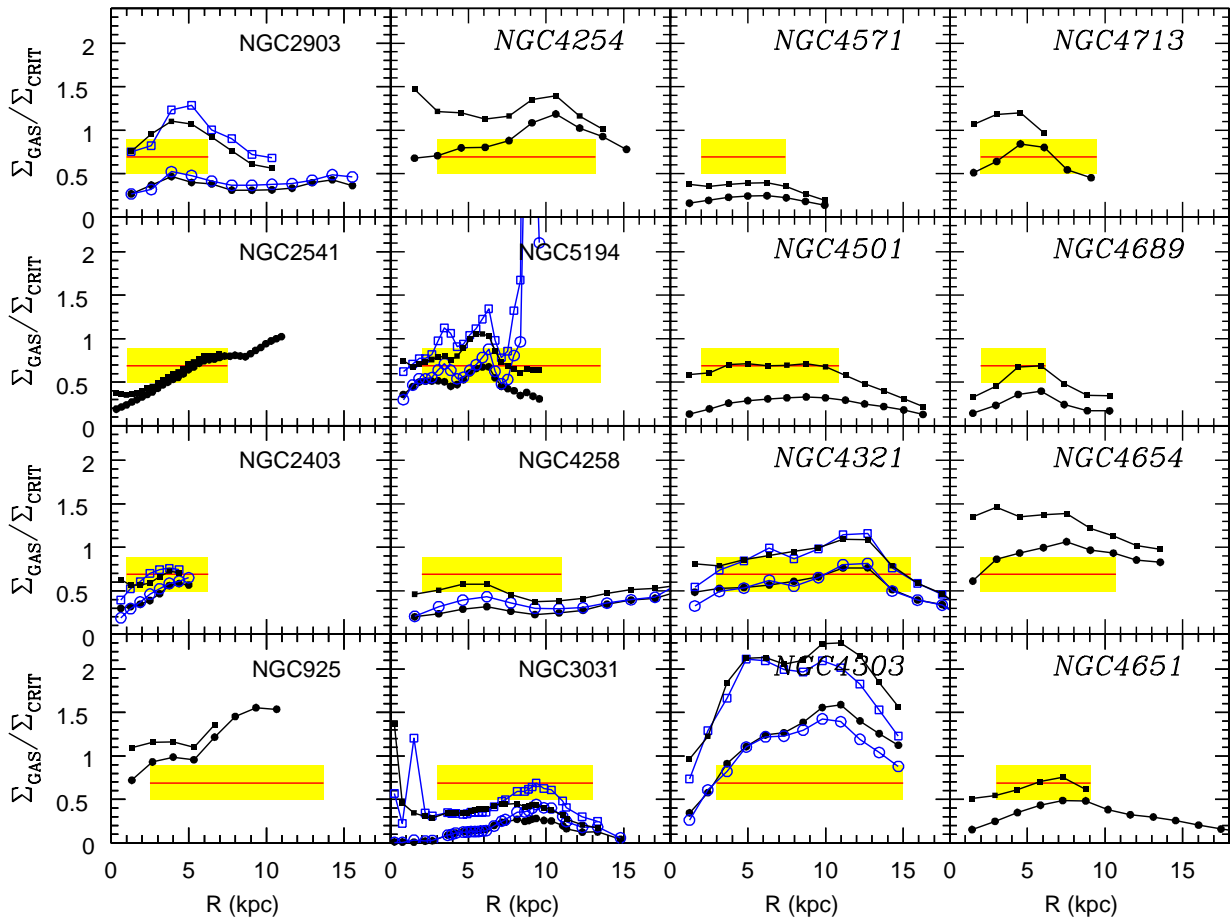} 
\caption{\label{figcritall}
Ratio of the gas density to the critical density, when gas only is
taken into account in the instability criterion ({\it filled circles}) and 
when the stellar density is taken into account ({\it filled squares});
the epicyclic frequency  was computed with the analytical rotation curve of
Equ.  \ref{pseudorota} (section \ref{secrotacurve}).
When a high resolution rotation curve was available \citep{sofue99}, the
epicyclic frequency was numerically derived and the corresponding results
are indicated by {\it open circles} and {\it squares}, respectively.
The horizontal line and the shaded area at 0.69$\pm$0.2 correspond to the threshold
values for star formation
found by Martin \& Kennicutt (2001); the inner limit of that
line corresponds to the bulge/disk limit and the outer one to the
observed threshold in star formation (see Sec. 2.4).
Galaxies with names in {\it italics} belong to the Virgo cluster.} 
\end{figure*}

\subsection {Thresholds for SF in galactic disks?}

Figure \ref{figcritall} shows the profiles of the ratio
$\Sigma_{GAS}/\Sigma_{CRIT}$ for the galaxies for which we have all
the required information. The ratio was computed through equations
\ref{eqcrit} ({circles}) and \ref{eqcritaeff} ({squares}). The
epicyclic frequency was computed by adopting the analytical rotation
curve of Equ. \ref{pseudorota} ({filled symbols}), which was derived
analytically.  For seven galaxies, the high resolution rotation curves
of
\citet{sofue99} were also used and numerically derived,
with very similar results (see Fig. \ref{figcritall}, {open
symbols}); an exception was the case of NGC 5194, in reason of its
peculiar rotation curve.  The absence of difference between the two
computations shows that the details of the rotation curve affect only
little the critical density, compared to other quantities.
 
When the critical density is computed with only the gaseous component
(Equ. \ref{eqcrit}), the disks are often found to be sub-critical
over large intervals of their radial extent. Eight of the galaxies of
our sample (2903, 2403, 4258, 3031, 4571, 4501, 4654, 4651) are found
to be subcritical with respect to the average threshold value of
$\Sigma_{GAS}/\Sigma_{CRIT}$=0.69$\pm$0.20 found by Martin and
Kennicutt (2001); the gas density is substantially lower than the
critical value even in regions with intense observed star formation.
The rest of our sample is equally divided between critical (2541,
5194, 4321, 4713) and over-critical galaxies (925, 4254, 4303, 4654);
in the latter case the value of $\Sigma_{GAS}/\Sigma_{CRIT}$ is
considerably larger than the threshold value of Martin \& Kennicutt
(2001).  We note that all three cases are encountered with similar
frequencies in isolated and Virgo spirals, i.e. the environment does
not seem to play a role in the overall disk instability.

As in the case of the Milky Way, including the stellar component in
the instability criterion (Equ. \ref{eqcritaeff}) leads to larger
values of the ratio $\Sigma_{GAS}/\Sigma_{CRIT}$ in the whole
disk. Only two disks still remain sub-critical with the modified
criterion (3031, 4571) and only marginally so.  In a few cases (4303,
4654) the modified values of $\Sigma_{GAS}/\Sigma_{CRIT}$ are three
times larger than the threshold value found by Martin \& Kennicutt
(2001).

\begin{figure*}
\centering
\includegraphics[angle=-90,width=\textwidth]{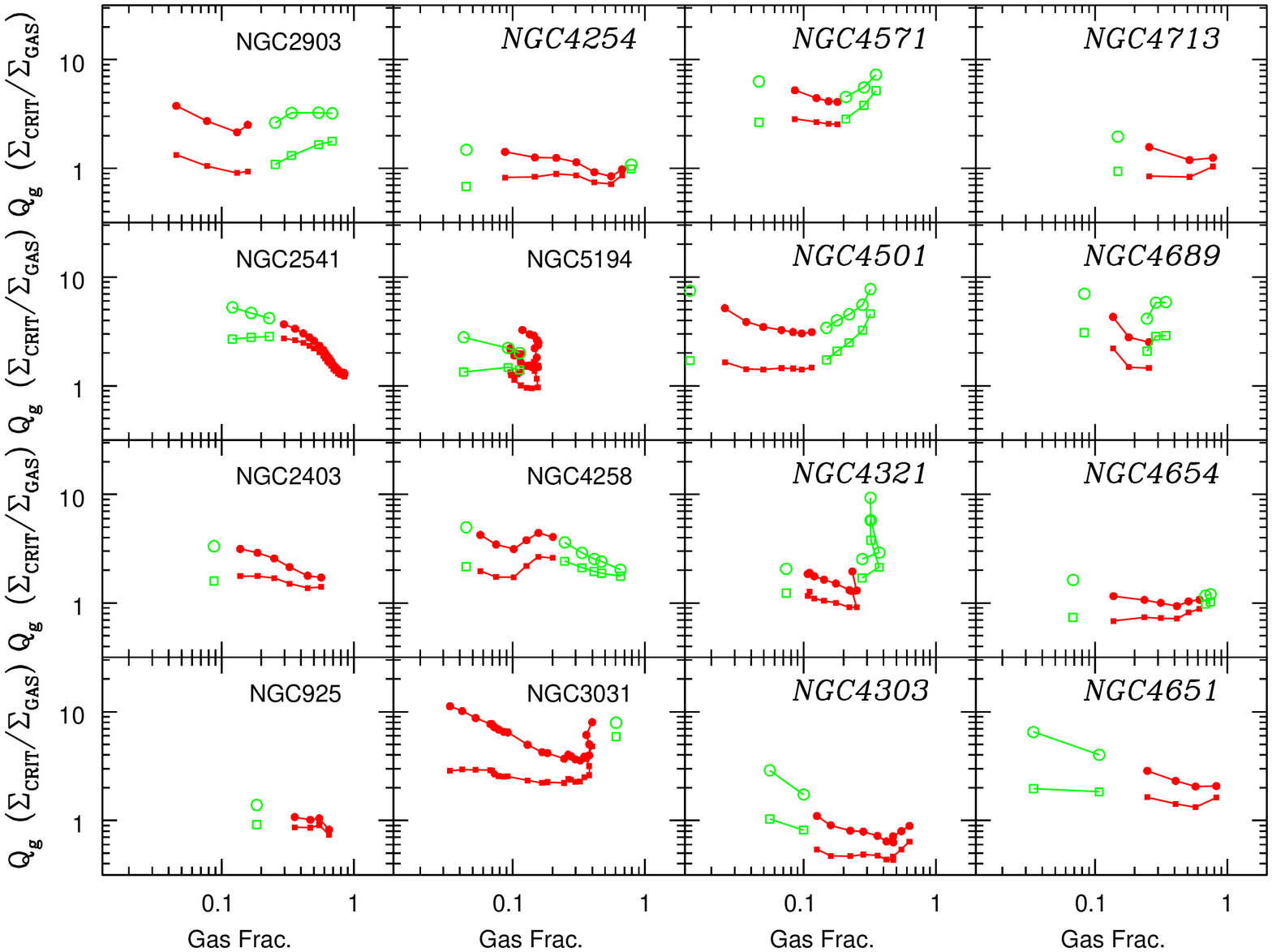} 
\caption{\label{figcritall2}
Ratio of the gas density to the critical density  as a function of
the gas fraction for all the disks of our sample, when gas only is
taken into account in the instability criterion ({\it circles}) and 
when the stellar density is taken into account ({\it squares}).
{\it Filled symbols} correspond to regions of active star formation
(i.e. inside the shaded bars of Fig. 8) and {\it open symbols}
to regions outside. 
}
\end{figure*}

The average value of $\Sigma_{GAS}/\Sigma_{CRIT}$ at the outer edge of the
star forming disks in our sample is
0.50 when only the gaseous component is included in the instability
criterion and 0.73 when the stellar component is also included. 
Due to the large variations between the values of 
$\Sigma_{GAS}/\Sigma_{CRIT}$ in our disks, we find a large dispersion
in the average threshold values : $\pm$0.33 in the former case and $\pm$0.38
in the latter, i.e. almost the double of the dispersion found by
Martin \& Kennicutt (2001).

We found that the evaluation of the H$_2$ profile in the outer parts
of the galaxies is crucial for the derivation of \Str. When the H$_2$ surface
density profile is not extrapolated beyond the last observed point
(i.e. assuming that no molecular gas is present beyond that point), 
we obtain a threshold value \Str=0.32 $\pm$ 0.16.  When a fit is performed to the
H$_2$ profile, and extrapolated beyond the last observed point, the result is
the previously mentioned 0.50
$\pm$ 0.33; this is in agreement with the value of 0.69 $\pm$ 0.2 of
\citet{martin01} who used the same assumption. 
 
In conclusion, contrary to the case of the Milky Way disk,
our analysis of external spirals does not
find evidence supporting the application of the (original or modified)
Toomre criterion for large scale star formation; it is true, however,
that the inclusion of the stellar component in the instability criterion
makes most of our disks overcritical according to that criterion. 
Also, it is worth noting that stars
are very often present beyond the threshold radius, as it can be seen
on the photometric profiles (Fig. \ref{figprofphot}), suggesting that
some star formation must have occurred in those regions in the past.

Our conclusions differ from the ones of Martin \& Kennicutt (2001)
despite the fact that we have a large number of galaxies in common 
(13 out of 16 in our sample).
In their larger sample of 32 disks they find only seven sub-critical
(i.e. 22 \%), whereas we find that eight of our 16 disks are sub-critical.
Among the galaxies in common, four were found to be subcritical in
our study but not in Martin \& Kennicutt (2001): NGC 2903, 4501, 
4651, 4689. Inspection of figure \ref{figprofobs} shows that the
molecular gas represents a significant fraction of the total amount
of gas over a large fraction of the disc within the threshold
radius for these galaxies. At the same time, 
Fig.  \ref{figCOfact} indicates
a quite large effect of the metallicity on the determination
of the H$_2$ surface density: with a constant conversion factor,
Martin \& Kennicutt overestimate the amount of H$_2$ with
respect to us.
The differences in the fraction of sub-critical disks and 
in the value of \Str \
in our study and theirs mainly reflect 
differences in the evaluation of the conversion from $CO$ to H$_2$.

We note that recently Wong \& Blitz (2002) reached
similar conclusions with us, concerning the validity of the
Toomre criterion. Their analysis of a different sample of galaxies
failed to show a clear relationship between the $Q$ value and
large scale star formation. They note, however, that $Q\sim$1 is often
observed in galactic disks, while two disks with high $Q$ values (sub-critical)
have small gas fractions; they suggest then that {\it Q is actually a measure
of the gas fraction in the disk} and they support their suggestion with a
rather qualitative analysis (see their Sec. 5.3).

We have used our detailed stellar and gas profiles to derive
corresponding gas fraction profiles, computed as
$\Sigma_{GAS}/(\Sigma_{GAS}+\Sigma_{*})$, for all the galaxies of our
sample and we plotted the $Q$ values as a function of the gas fraction
(Fig. 9). 
If \citet{wong02} were right, we would expect to find an
anti-correlation between Q and the gas fraction, large values of Q
corresponding to low gas fractions. On the contrary,
our results suggest no correlation between low values of the
gas fraction and sub-critical disks: Nine of our disks are indeed
sub-critical at a gas fraction of $\sim$0.1, as argued by Wong and
Blitz (2001), but seven are clearly critical. Also, we have several
galaxies which are sub-critical at gas fractions $\sim$1, whereas they
should be critical according to Wong \& Blitz (2001).  Thus, our
analysis does not support the idea that $Q$ is a measure of the gas
fraction in the disk.

\section{The star formation  law}

\label{seclaw}

\subsection{Star formation vs gas amount}

   \begin{figure*} \centering
   \includegraphics[angle=-90,width=\textwidth]{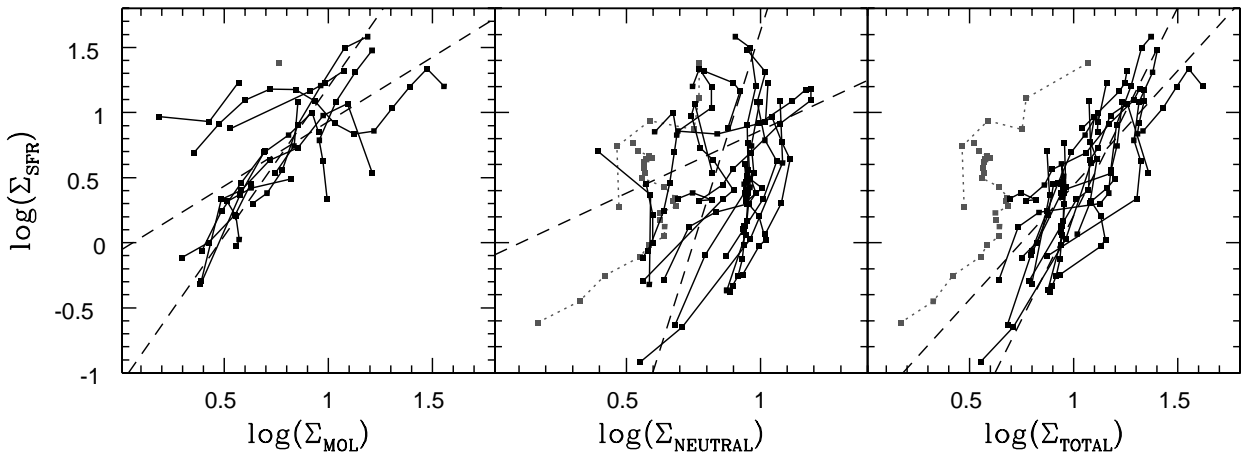} 
   \vspace{-6cm} 
    \caption{Star formation rate density as a function of
   molecular gas density ({\it left}), neutral gas density ({\it middle})
   and total
   gas density ({\it right}); units are ${\rm M_{\odot} pc^{-2} Gyr^{-1}}$ for
   the SFR and ${\rm M_{\odot} pc^{-2}}$ for the gas (where
   the He fraction has been accounted for).  The data are smoothed to the HI
   resolution. The {\it solid curves} connect points within the same  galaxy.  The
    {\it dashed lines} are the two regression fits in each case (see text);
   the galaxies NGC3031 and NGC4258 (indicated by the {\it dotted curves}
   are excluded from the fit.
   \label{gassfr}}
\end{figure*}

A comprehensive review of our current (non) understanding of large scale star
formation in galaxies has been recently presented by Elmegreen (2002), who also
provides an extensive reference list on the subject.

Intuitively, star formation should be associated to the molecular
content of a galaxy, rather than to the neutral gas amount. Such a
correlation has indeed been claimed to be observed in spirals by Young
and collaborators
\citep[e.g.][]{rownd99}, under the assumption of a constant conversion factor of CO to H$_2$.
Similar conclusions were reached by \citet{boselli95}, who
showed that the molecular mass
follows quite closely the star formation rate in luminous
spirals; the galaxies of their sample display similar metallicities and the
assumption of a constant conversion factor is then justified.
However,  it was argued by \citet{boselli95} that 
the scatter obtained for less luminous
galaxies could be due to variations in the conversion factor.
\citet{boselli02b} showed recently that the conversion
factor depends indeed on the physical properties of the galaxy. 
Once a metallicity-dependent conversion factor is adopted, a good
correlation is observed between H$_2$ and the star formation rate.

As discussed in Sec. 2.3 we adopted in this paper a 
metallicity-dependent conversion factor to compute the densities 
of H$_2$ within the disks of our sample. 
It can be seen on Fig. \ref{gassfr} that the resulting azimuthally averaged
molecular gas surface density presents a better correlation with the  
star formation rate density than the neutral gas 
(left and middle panels, where
the H$_2$ and SFR profiles have been smoothed to the HI resolution, 
and corrected for the He fraction by multiplying them by 1.4).  
The good agreement correlation between H$_2$ and the star formation rate
in our analysis of azimuthally averaged quantities corroborates
the findings of \citet{boselli02b} concerning global quantities of the disks.

Instead of thinking the star formation process as a sequence (neutral
gas towards molecular gas towards stars), one may consider it as the
result of dynamical processes involving the totality of 
the gas density. In that case, 
a better correlation should be found between the star formation rate
and the total gas density rather than either the molecular or the
neutral component. Indeed, 
since \citet{schmidt59}, there has been evidence that the 
star formation rate density ($\Sigma_{SFR}$) 
is related to the total gas density ($\Sigma_{GAS}$) 
 \citep[e.g.][]{kennicutt98}:

\begin{equation}
\label{eqsfr1}
\Sigma_{SFR} = \alpha \Sigma_{GAS}^{n} \; (1<n<2). 
\end{equation}

Basic models based on self-gravitating discs can produce such a ``Schmidt
law'' \citep[e.g.][]{kennicutt98b}.  The right panel of Fig. 
\ref{gassfr} displays a reasonably good correlation
between the star formation rate and the gas density
for most of the galaxies of our sample. Two
galaxies which are rather peculiar can be easily identified on
this plot.  One of them is NGC 3031 which has a low gas content but a
high star formation rate, especially in its inner part;
this galaxy is forming stars despite being sub-critical (Fig. 9). 
The other is NGC 4258 and displays
a normal gas content but a larger efficiency to form stars than all
the other galaxies in our sample.

We stress that the correlations displayed in Fig. 10 concern {\it 
azimuthally averaged }
quantities, whereas Boselli et al. (2002) or \citet{kennicutt98}
studied such correlations between global quantities or averaged
over the whole disk (inside the optical radius).

The Schmidt law is  the most widely used, but other semi-empirical
``recipes'' of star formation have been suggested for galactic disks
(and implemented in relevant models).

Ohnisi (1975)  suggested 
that star formation is enhanced by the 
passage of spiral density waves with a frequency 
$\Omega(R)-\Omega_P$, where $\Omega(R)$ is the galaxy's rotational frequency
at radius R and $\Omega_P$ the frequency of the spiral pattern; for
$\Omega(R)>>\Omega_P$ this leads to a star formation rate
SFR $\propto\Omega(R)\propto V(R)/R$, where V(R) is the corresponding
rotational velocity \citep[see also][]{wyse89}. According to \citet{larson88}
the presence of the $\Omega(R)$ factor in the radial variation
of the star formation rate could also be obtained if
a self-regulation of star formation occurs such that
$\Sigma_{GAS} \sim \Sigma_{crit}$. Kennicutt (1998a) has
indeed found that the observed averaged (over the optical radius)
densities of the star formation rate and the total gas density are related
equally well either by a simple Schmidt law (Equ. 9) or by
the form:
\begin{equation}
\label{eqsfr2a}
\Sigma_{SFR} = \alpha \Sigma_{GAS}/\tau_{DYN}
\end{equation}
where $\tau_{DYN}\sim R/V(R)$ is the dynamical timescale
at radius $R=R_{OPT}$.

In \citet{boissier99}, a ``semi-theoretical'' SFR deduced from 
those consideration was adopted:
\begin{equation}
\label{eqsfr2}
\Sigma_{SFR}=\alpha \Sigma_{GAS}^{n} V/R,
\end{equation}
($V$ being the rotational velocity, taken constant and equal to 220 km s$^{-1}$ for the 
Milky Way). The index $n$ was chosen to be $n$=1.5 on empirical
basis (providing the observed relation between $\Sigma_{SFR}(R)$ and $\Sigma_{GAS}(R)$ 
and reproducing the observed abundance gradient).
This theoretical star formation rate was adopted in subsequent models
extended to all spiral galaxies, which reproduced successfully their global
properties \citep{boissier2000,bo2pg} and their abundance and colour
gradients \citep{prantzos2000}.

Another star formation law, sometimes used in models of chemical evolution
\citep[e.g.][]{matteucci01} was 
suggested by \citet{dopita94} on the basis
of an observed correlation between the stellar surface density (as traced
by I-band photometry) and the star formation rate:
\begin{equation}
\Sigma_{SFR}=\alpha \Sigma_{GAS}^{n} \Sigma_{T}^{m}
\label{eqsfr3}
\end{equation}
where $\Sigma_T$ is the total surface density (gas plus stars).
\citet{dopita94} tested only indirectly this law (via a model,
since they had no gas data available) and found it to be consistent
with observations if $n+m$ is between 1.5 and 2.5. They also presented 
a model of stochastic self-regulating star formation, 
predicting $n$=5/3 and $m$=1/3, values that are in agreement with
their empirical findings.


\subsection{Empirical determination of SF parameters}
\label{seclawfits}

We tested the three SF laws of equations \ref{eqsfr1}, \ref{eqsfr2} and
\ref{eqsfr3} with our data (with all the data being smoothed to the HI
resolution).  The gas profile was multiplied by a factor 1.4 to take
into account the He contribution.  The H$_2$ data were extrapolated
assuming an exponential profile (see below  for a discussion
of this assumption).  Equ. \ref{eqsfr2} requires the
rotational velocity $V(R)$, that we simply approximated by the
analytical rotation curve of Equ. \ref{pseudorota}.
For Equ. \ref{eqsfr3} we need an estimate of the total
surface density, that is of the stellar density; the
photometric H band profiles, barely affected by dust extinction,
were used for that purpose, assuming the same mass to luminosity 
ratio ($M_*/L_H$) as in \citet{bo2pg}.

\begin{figure}
\centering \includegraphics[angle=0,width=0.5\textwidth]
{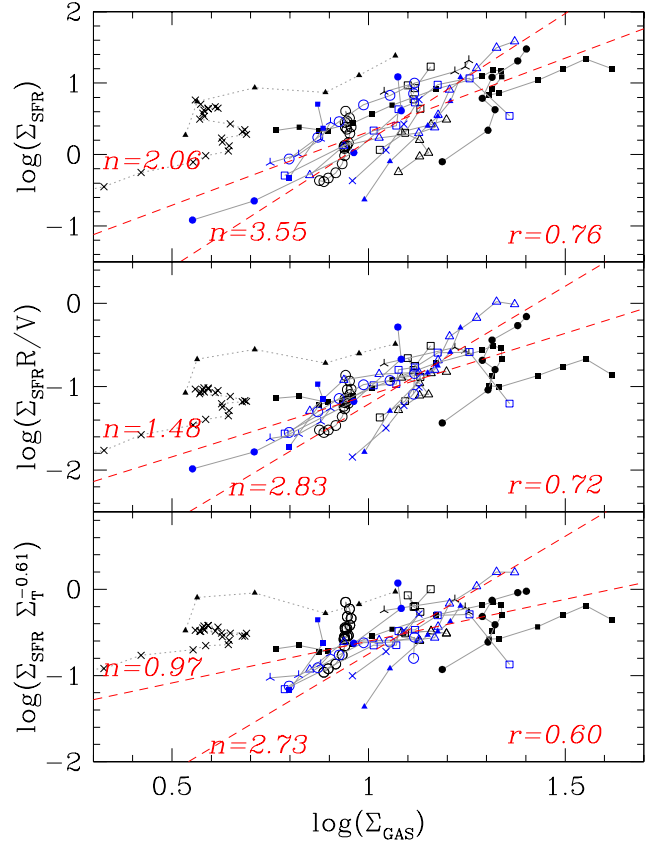} 
\caption{ Test of various star formation laws.  {\it Top}: gas surface
density vs Star Formation Rate surface density (testing the standard
Schmidt law, Equ. \ref{eqsfr1}). {\it Middle}: gas density vs
$\Sigma_{SFR} \times R / V(R)$ (testing Equ.
\ref{eqsfr2}). {\it Bottom}: gas density vs $\Sigma_{GAS}^n
\Sigma_T^{-0.61}$ (testing a Dopita- Ryder law, Equ. \ref{eqsfr3}, with
the best value of the second parameter $m$=0.61).  
The {\it grey lines} connect points of the same
galaxy (the {\it dotted lines} correspond to NGC4258 and NGC3031, not
considered in the fits, see section \ref{secngc4258}).  
Different symbols are used for different galaxies.
The {\it dashed
lines} show two fits, one minimizing deviations on $\Sigma_{SFR}$ (the
one that should be used for a determination of the SFR parameters) and one
minimizing deviations on $\Sigma_{GAS}$. The corresponding index
$n$ is indicated for each one of them and is also reported on 
Table \ref{tablefits}. The correlation coefficient (r) is indicated in the
right-bottom part of each panel. 
\label{gassfr2}}
\end{figure}

Fig. \ref{gassfr2} shows our data plotted  in the plane
$\Sigma_{GAS}$ vs $\Sigma_{SFR}$ ({\it top}, testing  Equ. 9),
in the plane $\Sigma_{GAS}$ vs $\Sigma_{SFR} \times R / V(R)$ 
({\it middle}, testing Equ. \ref{eqsfr2}) and in
the plane $\Sigma_{GAS}$ vs $\Sigma_{SFR} \times \Sigma_{GAS}^n
\Sigma_T^{-0.61}$ ({\it bottom}, testing Equ. \ref{eqsfr3}).  
In the last case, the best-fit value for the
second parameter $m$ was adopted for the figure
(i.e. the one minimising the $\chi^2$ for the SFR).

The points corresponding to a single galaxy are connected by a grey line.
The dashed lines show simple least square fits, one minimizing the deviation on
the SFR density, and one on the gas surface density (the difference between the 
indexes in each case gives an idea of the dispersion of the relation).
Since we are searching for the ``best fit law'' of the SFR, we will adopt
for the discussion the first of these two fits.

The dotted lines correspond to NGC3031 and NGC4258. The latter 
is obviously an outlier, having
quite low gas density but relatively high SFR (see Fig. \ref{figprofobs})
and we do not consider it in the fits;
\label{secngc4258}
note that high resolution CO data for this galaxy differs from the low
resolution ones (see section \ref{secco2h2}).
We also ignored NGC3031,  because its 
gaseous profile (HI and total) is very atypical (see Fig. \ref{figprofobs}).
It is the only object in our sample with a low gas density over the whole disc.
NGC3031 is indeed among the more sub-critical discs of \citet{martin01}.

One would expect that the outlying or peculiar galaxies in Fig. 11
(as well as in Figs. 8,9,10) could be systematically associated to the
cluster environment, HI-deficient objects, or have a peculiar
morphology. We do not find obvious trends in our small sample, however
the analysis of Gavazzi et al. (2002) on integrated quantities has
shown that cluster HI-deficient galaxies have, on average, lower star
formation activities than their field counterparts.

For a simple Schmidt law (Equ. \ref{eqsfr1}), the least-square
fit gives an index $n$ $\sim$ 2, steeper than the 1.4 suggested by 
\citet{kennicutt98b} or the 1.3 $\pm$ 0.3 of his earlier work (Kennicutt, 1989). 
When a dynamical factor is introduced
(Equ. \ref{eqsfr2}), we obtain $n$ $\sim$ 1.5, steeper again 
than the slope $n \sim 1$ found by
\citet{kennicutt98b}. However the introduction of $\Omega$ in
both studies acts in the same direction: it reduces the value of $n$.

The differences between our results and those of 
\citet{kennicutt98b} may be due to several reasons:
i) \citet{kennicutt98b} uses surface densities averaged within the
optical radius, i.e. he includes the bulge region (while we only study
the star forming disk) and in some 
cases he also includes some gas located beyond the threshold radius
(while we work on radial profiles within the radial range  of interest);
ii) The law obtained by \citet{kennicutt98b} was established on a 
large range of gas surface density (an astonishing factor 10$^5$)
 by introducing in the sample star-burst galaxies; however, 
for normal discs only, his analysis also showed a large dispersion
of the index $n$, with values ranging 
from 1.29 for a standard least square to 2.47 for a 
bivariate least-square regression (for a standard Schmidt law,
Equ. \ref{eqsfr1}). The corresponding values in our case are
respectively 2. and 2.6.
iii) We use a conversion factor between CO and H$_2$ that depends explicitly
on the metallicity, while he adopts a constant conversion factor. This important
point also applies to the comparison of our work and the one of Kennicutt (1989),
also based on radial profiles of spiral galaxies.
iv) The galaxy sample  is different; it is obvious from Fig.
\ref{gassfr2} that star formation in galaxies  does not obey a well
defined relation but
present a large scatter around an average one. The 
SFR law can then be defined  only in a statistical sense, and since the
number of galaxies for which such studies can be done is still small, it
is understandable that studies with different samples produce different
results.

\citet{wong02} used a small sample of 7 galaxies only, but with
higher spatial resolution. 
Assuming a uniform  model of extinction (comparable to the present work) 
they found for the simple Schmidt law a value of $n$=1.1, i.e.
they found a relation less steep than both us and \citet{kennicutt98b}.
On the other hand,  assuming a dependence of extinction on the gas column density
they found $n$=1.7. They also found that a SFR of the form $\Sigma_{GAS} 
\Omega$ is compatible with their data, but only if 
an important radially dependent extinction is assumed. 
However, they did not attempt to fit a law of the form of Equ. \ref{eqsfr2}.

The differences between our work and \citet{wong02} may result from:
i) The small statistics (as previously discussed);
ii) The fact that the galaxies of \citet{wong02} are 
selected to be bright in CO, i.e. H$_2$-rich, which may introduce some bias
in the sample, as these authors also acknowledge;
iii) The sensitivity of the obtained index $n$ to the assumed 
extinction gradient, which is difficult to evaluate since
it depends also  on the abundance gradient, and since the observed
scatter at a given radius is stronger than that gradient \citep{martin01}.
iv) The assumption of a constant conversion factor of
CO to H$_2$.

Finally, we tested the SF law proposed by Dopita \& Ryder (Equ. \ref{eqsfr3})
with the sub-sample for which the stellar density profile is available from 
H-band photometry. A simple least square fit
(minimizing the deviation on $\Sigma_{SFR}$) to our data leads to
$n \sim$ 1 and $m \sim$0.61, substantially  different from the 
values 5/3 and 1/3   obtained by \citet{dopita94}.

The results presented in Fig. \ref{gassfr2} are obtained by 
adopting the analytical rotation curve of
Equ.  \ref{pseudorota} and an extrapolation of the molecular gas profile,
as described in Sec. \ref{secthres}. To test the effect of these assumptions,
we did the same work assuming that there is no molecular gas beyond the last
CO detection point. The difference in the slopes of the relationships
are presented in Table \ref{tablefits}.
\begin{table}
\caption[]{\label{tablefits}Slopes of the star formation laws under two 
different assumptions on the
molecular content at large radii. 
The values of the exponent $n$ are given for the two regression lines in
each case. For the Dopita-Ryder law, the ``best-fit'' value of the second 
parameter ($m$) was assumed.
Note that adopting a flat rotation curve barely affects 
the values of $n$ for the
$\Sigma_{GAS}^n \Omega$ law.  }
\begin{tabular}{l l l }
\hline
\hline
$\Sigma_{SFR}$                & No H$_2$ extrapolation & H$_2$ extrapolation \\
\hline
$\alpha \Sigma_{GAS}^n$              & 1.96 (3.1)             & 2.06 (3.55) \\
$\alpha \Sigma_{GAS}^n \Omega$       & 1.38 (2.52)	      & 1.48 (2.83) \\    
$\alpha \Sigma_{GAS}^n \Sigma_{T}^m$ & 1.00 (2.21) for m=0.56  & 0.97 (2.73) for m=0.61 \\ 
\hline
\end{tabular}
\end{table}

The differences are relatively modest, contrary to what occurred for the
threshold, since the fits are essentially determined by the
data concerning the main body of the molecular disk.

Adopting a flat rotation curve instead of the analytical
rotation curve of Equ. \ref{pseudorota} leads also to small differences: we obtain a
slightly lower slope (1.28 in the case of no H$_2$ extrapolation and
1.32 in the case of H$_2$ extrapolation) for a flat rotation curve
than with  the analytical rotation curve (respectively 1.38 and 1.48).

In conclusion, the SF laws that we obtained  depend only weakly on
assumptions concerning  the rotation curve and the 
molecular gas profile in the outer disk.

\subsection{Application to the Milky Way}

In the previous section, we determined the parameters of 
three widely used star formation laws (equations \ref{eqsfr1},
\ref{eqsfr2},\ref{eqsfr3}) by means of our galaxy sample. 
As in the case of the SF threshold (see Sec. 3.2), it is interesting to see how
the resulting SF laws match the star formation data of the Milky Way disk.

\begin{figure} \centering
\includegraphics[angle=-90,width=0.5\textwidth]{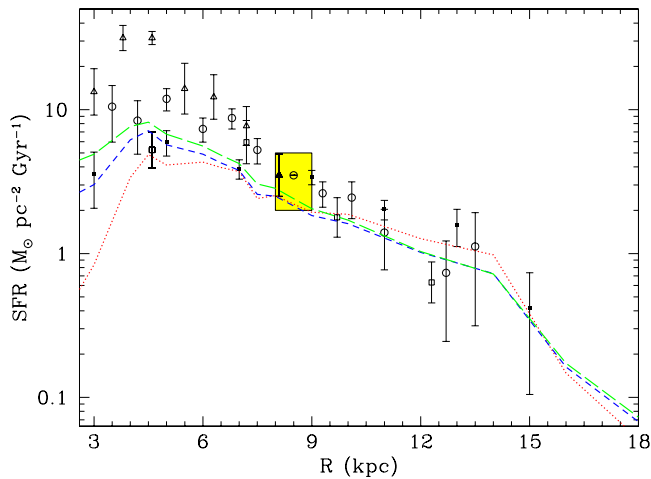}
\caption{\label{profilmw} Application of the three SF
laws obtained in section \ref{seclawfits} 
(pure Schmidt: {\it dotted}, Schmidt modified by rotation: {\it short-dashed},
and Schmidt modified by stellar density: {\it long dashed}). They are
compared with various  SF tracers in the disk of the
Galaxy (see section 4.3 for References).}
\end{figure}

Using the well known gaseous and stellar radial profiles of the Milky Way
and a flat rotation curve (V(R)=220 km s$^{-1}$), we
calculated the corresponding SF profiles; the results are displayed 
in Fig. \ref{profilmw}, respectively
for the simple Schmidt law ({\it dotted curve}), 
the Schmidt law modified by the dynamical factor V/R 
({\it short-dashed curve}) and the Schmidt law modified by the stellar density
({\it long-dashed curve}). 
We compare the results to the SF radial profile of the
Galaxy, as estimated through various tracers  
(\citet{lyne85}, 
\citet{gusten83}, 
\citet{guibert78}, 
and \citet{case98}). 
Note that the observations were scaled to the value of the SFR in 
the solar neighborhood which has its own uncertainty; the box in the center
of the figure indicates the range of its possible values \citep{rana91}.

The three SF laws produce quite similar results in the range 4-18 kpc,
i.e. in the whole radial extent of the disk.  The pure Schmidt law
fails considerably in the inner disk (inside 3 kpc), since the low
amounts of gas there do not allow it to reproduce the relatively high
values of observed SFR.  All three laws give satisfactory results in
the 4-15 kpc range, while the modified Schmidt laws fit the
observations also in the inner disk; however, due to several
complications that might arise in that region (i.e. role of the bar in
the star formation) the ``success'' of the modified Schmidt laws in
that region should not be considered as a ``proof'' a their validity.

In summary, data for the Milky Way disk are compatible with all three
SF laws that we empirically determined in the previous section.  Only
a consistent application of those SF laws in a chemical evolution
model of the Milky Way could (perhaps) favour one of these SF laws
over the others, by comparison to as many observed radial profiles as
possible (e. g. chemical and photometric profiles, as in the work of
Boissier \& Prantzos 1999 for the gaseous, chemical and photometric
profiles of the MW disk). We note, however, that in practice, even
this test may not be conclusive, because of the poorly known radial
and temporal dependence of the gas infall rate on the disk.  Gaseous
infall is indeed required, at least in the solar neighborhood to
account for the so-called ``G-dwarf problem'' (e.g Pagel 1997 and
references therein).

\section{Summary}

\label{conclu}

In this paper we present a detailed study of the  properties of star 
formation in disk galaxies. We use an extended data set, consisting of
(azimuthally averaged) radial profiles of stellar and gaseous
(atomic and molecular) surface densities, of star formation rates and
rotation curves. 
The molecular gas is derived from CO with a metallicity dependent 
conversion factor for the first time in this kind of study.
These data allow questions related to 
the  star formation rate and to the existence of  a star formation
threshold in galactic disks to be studied. 
Our sample comprises 16 disk galaxies, about
half of which belong to the Virgo cluster. 

Our results may be summarised as follows:

The existence of star formation threshold in our disks was studied by using 
the Toomre (1964) instability criterion and using either the gaseous
component alone or both the gaseous and stellar components in the definition
of the instability parameter Q. In the latter case, the profile of stellar
velocity dispersion is also required and we calculated it for our disks
based on the observations of Bottema (1993). We also used both analytical
and observed rotation curves for the evaluation of Q.

We find that only half of our sample galaxies are overcritical
(i.e. locally unstable and prone to form stars) when the gaseous component 
alone is taken into account, even in disk regions where observations indicate
that vigorous star formation is taking place.
Including the stellar component improves substantially the situation, since
only two disks remain sub-critical then (and only marginally so).

These results refer to an average value of the instability parameter 
Q=0.70$\pm$0.20 (i.e. slightly lower than unity), as observationally determined
by Martin \& Kennicutt (2001) with a larger disk sample (32 galaxies).
Our own Q value, determined at the termination of the SF profile, is similar
to the above value  when the stellar component is included and slightly 
lower when it is ignored (0.71 and 0.5, respectively). However, in both cases,
corresponding uncertainties are larger than found by 
Martin \& Kennicutt (0.38 and 0.33, respectively, compared to 0.20 in their 
case).

We find that the instability criteria apply fairly well to the Milky Way disk,
where Q values are systematically $\sim$0.7 when the gaseous component alone is
used and $\sim$0.9-1 when the stellar component is also included. We
find it encouraging that the galaxy with the best defined profiles shows
such an exemplary behaviour. We think that the large dispersion in the
Q values found in our sample result (at least partly) from systematic 
uncertainties in the various profiles entering the evaluation of Q.
Among them, of crucial importance are:
i) the conversion factor of CO emission to H$_2$ gas amount (we use a
metallicity dependent conversion factor, unlike previous studies) and,
even more so
ii) the extrapolation of the H$_2$ profile beyond the last observed
point. These factors explain to a large part the quantitative differences
between our results and those of  Martin \& Kennicutt (2001) and Wong
\& Blitz (2002). In particular, our findings do not support the claim
of Wong \& Blitz (2002) that Q is actually
a measure of the gas fraction in the disks.

In agreement with  Wong \& Blitz (2002) we find that the observed local
(i.e. at a given radius)
SF rate density correlates better with the total gas density or with the
molecular gas density than with the neutral gas density.
We analysed three local SF laws with our data: 
a pure Schmidt law  SFR$\propto\Sigma_{GAS}^n$, 
a Schmidt law modified by the disk rotational frequency 
SFR$\propto\Sigma_{GAS}^n V(R)/R$, and a Schmidt law modified by
the local surface density $\Sigma_{T}$:  
SFR$\propto\Sigma_{GAS}^n \Sigma_{T}^m$ (according to a suggestion
by Dopita \& Ryder 1994).

We find that the modified Schmidt laws do slightly better than the pure
Schmidt law, as expected in view of their supplementary degrees of
freedom. We also find a larger index $n$ for the gaseous component
than Kennicutt (1998), both for the pure Schmidt law and for the one
modified by rotation (2 and 1.5, compared to 1.4 and 1,
respectively). We note that Kennicutt's analysis included starburts,
so that his gas surface densities covered five decades in magnitude,
instead of $\sim$1.5 decade in our case; by limiting his analysis to
the low surface density normal spirals, Kennicutt (1998) also found
larger values of $n$, close to ours. Besides, his analysis concerned
only averaged quantities over the galactic disks, whereas ours
concerns azimuthally averaged quantities, a fact which certainly
explains the largest dispersion and the poorest fits that we
obtain. On the other hand, Wong \& Blitz (2002) used only azimuthally
averaged quantities in their study and found values of $n$=1.1 for a
uniform extinction model and 1.7 for an extinction dependent on the
column density in the case of the pure Schmidt law; the latter case
corresponds more to our own results.

Again, we find that the three derived SF laws apply fairly well to the
data of the Milky Way disk, although the pure Schmidt law fails in the
inner Galaxy (where the situation is uncertain anyway, due to the
poorly known role of the galactic bar). The exemplary behaviour of the
Milky Way disk makes us think that (either it is exceptional or) a
much more systematic work than this one could allow to determine much
better the local properties of star formation in disks. Such a work
should involve: a much larger number of (unperturbed) disks than used
here; much more extended and detailed radial profiles, especially in
the case of molecular gas (for instance the results of the recent BIMA
survey, Regan et al. 2001); and, above all, a much better
understanding of the various systematic uncertainties of the problem,
for instance concerning the absolute local values of the star
formation rate and the role of non axisymmetric profiles.

\def\mnras{MNRAS}
\def\aj{AJ}
\def\apj{ApJ}
\def\apjs{ApJS}
\def\aap{A\&A}
\def\aaps{A\&AS}

\end{document}